\documentclass[12pt]{article}
\usepackage{amssymb}
\usepackage{setspace}
\begin{document}

\newcommand{\bra}[1]{\langle #1|}
\newcommand{\ket}[1]{|#1\rangle}
\newcommand{\braket}[2]{\langle #1|#2\rangle}
\newcommand{\en}{\textrm{End}}

\def\maketitle{
   \begin{flushright}
   J. Math. Phys. \textbf{45}, 2492 (2004)\\
   DAMTP-2003-93
   \end{flushright}
   \begin{center}
   \Huge{Superconformal Primary Fields on a Graded Riemann Sphere}\\
   \vskip 0.8 cm
   \normalsize{Jasbir Nagi}\\ \vskip 0.2 cm \normalsize{J.S.Nagi@damtp.cam.ac.uk}\\ \vskip 0.2 cm \normalsize{DAMTP,
   University of Cambridge, Wilberforce Road,}\\ \vskip 0.2 cm \normalsize{Cambridge,
    UK, CB3 0WA}
   \vskip 0.8 cm
   \end{center}
   }

\maketitle
\begin{abstract}
Primary superfields for a two dimensional Euclidean superconformal field theory are
constructed as sections of a sheaf over a graded Riemann sphere. The
transformation law is found to be the same as that of an $O(N)$
extended primary field. The
construction is then applied to the $N=3$ Neveu-Schwarz case. Various
quantities in the $N=3$ theory are calculated, such as
elements of the super-M\"obius group, and the two-point
function. Applications of the construction to calculate three-point
functions and fusion rules in a manifestly supersymmetric fashion are discussed.
\end{abstract}

\newpage
\tableofcontents

\section{Introduction}
Two-dimensional conformal field theory has many applications in
statistical mechanics and string theory. It also has a very rich algebraic
nature, in a sense due to the symmetry algebra being
infinite-dimensional. Exactly how this works in the bosonic case is
extremely well understood.

The supersymmetrisation of two-dimensional conformal field theory is
something that has been studied heavily from a string theory
approach. The string is described by a two-dimensional conformal field
theory, and the supersymmetrisation of the conformal field theory
essentially admits fermions onto the string. This has mostly been
studied from a Lagrangian point of view, where the Lagrangian exhibits
the classical symmetries. Canonical quantisation can then be used to then
obtain the quantum algebra.

In bosonic conformal field theory there is a way, using the high
degree of symmetry, to obtain the quantum algebra by algebraic means,
rather than from a Lagrangian. Extending this to a theory with one or
two Grassmann variables has been covered extensively in the
literature. Adding more supersymmetry to the theory has been studied,
e.g.\cite{ks} \cite{ck} \cite{cohn}, but in nowhere near as much depth
as the $N=1,2$ theories.

In two dimensional conformal field theory, it is found that one can
always find a conformal transformation that maps the two-dimensional
theory to a two dimensional theory that is flat. In the Euclidean
case, the conformal transformations that map the plane to itself are
precisely the holomorphic and anti-holomorphic transformations of the
plane to itself. One can therefore build Euclidean two-dimensional
conformal field theory on a complex plane. To get a more `global'
picture of what is going on, the theory can then be conformally mapped
to the Riemann sphere. Many of the properties of the conformal field
theory can then be described by the properties of the Riemann sphere. The
question then becomes how to build a
theory on a `super Riemann sphere', and how the properties of this
object can be related to the properties of a superconformal field
theory. In this note, the question is
addressed, with particular attention payed to the $N=3$ case.

Superconformal algebras in the classical case look like derivations on
a polynomial ring in $\{ z, z^{-1}, \theta_i \}$, where the $\theta_i$ are
anticommuting `co-ordinates', that preserve a
differential form. This can be combined with the theory of extended graded
manifolds, with the manifold in question being the Riemann sphere, to
give a suitable setting for superconformal field theory (sections
2-4). One can then use this setting to construct and calculate various
quantities in the field theory. The $N=3$ Neveu-Schwarz case is studied in detail
(sections 5-10).

\section{The Graded Riemann Sphere}
In this section, a Riemann sphere is considered, and how one can generalise
it to the superconformal case. It should be mentioned that there are
strictly speaking two approaches, which give rise to the same
structure \cite{rog3}, \cite{bat2}. Here, the algebraic structure\cite{kos},
known as a Graded Manifold (of the extended type) will mostly be used. In some
instances, it will become necessary to transfer to the analytic point
of view, namely Supermanifolds \cite{rog1}.

First consider an ordinary Riemann sphere. Rather than consider it as
a geometric object, one could consider it as a collection of open
sets, $U_i$, and consider the functions $f:U_i\rightarrow\mathbb{C}$ that are holomorphic in each
open set, denoted $f(U_i)$. Each $f(U_i)$ is a ring under
addition in $\mathbb{C}$, and pointwise multiplication. Given an open
subset $V\subseteq U$, one can construct a non-singular ring homomorphism $\rho_{UV} :
f(U)\rightarrow f(V)$. These ring homomorphisms become the fundamental
tools to work with. They give a way of comparing $f(U_i)\mid_{U_i\cap U_j}$ and $f(U_j)\mid_{U_i\cap U_j}$ and allow one to construct a sheaf of rings over
the Riemann sphere, denoted $\mathcal{A}_0$
\cite{kos},\cite{bred}. One can see that each $f(U_i)$ will be a
subring of $\mathbb{C}\lbrack z, z^{-1} \rbrack$. One can then
consider derivations on each $f(U_i)$, denoted Der$(U_i)$ which are $\mathbb{C}$-linear
maps from $f(U_i)$ to $f(U_i)$ that obey the Leibniz rule. Der$(U_i)$
then forms a rank one module over $f(U_i)$. The map $\rho_{UV}$
induces a map $\rho_{UV*}^{-1} : \textrm{Der}(U) \rightarrow
\textrm{Der}(V)$. One can use these $\rho_{UV*}$ to construct a sheaf
of abelian groups, namely the tangent sheaf, denoted
$\mathcal{D}^1\mathcal{A}_0$. It is, locally, a rank one
$\mathcal{A}_0$-module. A section of it can be written locally as
$g(z)\frac{\partial}{\partial z}$. On each $U_i$, one can also
consider the $\mathcal{A}_0$ linear maps of $Der(U_i)$ into
$\mathcal{A}_0$, denoted $\Omega^1(U_i)$. This is also a rank one
$f(U_i)$-module, and the $\rho_{UV*}$ induce a map $\rho_{UV}^*
: \Omega^1(U) \rightarrow \Omega^1(V)$. Once again, the $\rho_{UV}^*$ can
be used to construct a sheaf, denoted $\mathcal{D}_1\mathcal{A}_0$. It
is locally a rank one $\mathcal{A}_0$-module. Locally, a section can
be written as $dz g(z)$. One can the define the conformal condition as
demanding that $\rho_{UV}:z' \mapsto z$, for $z', z$ local co-ordinates in
$U, V$ respectively, as having the property $\rho_{UV}^*dz' =
dz\kappa(z)$ for some $\kappa \in \mathcal{A}_0$. Given this construction, a basis of infinitesimal transformations can be written down,
namely $z' = z + az^{n+1}$ corresponding to a space of vector
fields, which gives rise to the Witt algebra. Phrasing the structure of a Riemann sphere in this
way gives the most natural generalization to a graded Riemann sphere.

Similarly, an extended graded manifold can be defined,
where each ring associated to each open set is no longer a subring of
$\mathbb{C}\lbrack z, z^{-1} \rbrack$, but a larger ring containing
Grassmann generators. The ring is no longer over $\mathbb{C}$, but
over a complex, finitely generated unital Grassmann algebra, $B_{L'}$. Each
ring associated to each open set is now a subring of 
\begin{eqnarray}\label{1Nring}
B_{L'} \lbrack z, z^{-1}, \theta_1, \ldots, \theta_N \rbrack
\end{eqnarray}
It is worth noting that this is a slightly more general requirement
than that of a graded manifold. In the graded manifold case, the ring
is often still taken to be over $\mathbb{C}$, and so would look like
\begin{eqnarray}\label{oldring}
\mathbb{C} \lbrack z, z^{-1}, \theta_1, \ldots, \theta_N \rbrack
\end{eqnarray}
which, as a ring, is contained in (\ref{1Nring}). This can be seen by
constructing a map $\pi : B_{L'}\mapsto \mathbb{C}$ by projecting onto
the unital element in $B_{L'}$. The map defines an important quantity, namely the `body'
of an element of a grassmann algebra. The
approach of looking at an algebra over $B_{L'}$ rather than $\mathbb{C}$, as far as the author is aware, was first introduced in
\cite{rog3}. The ring (\ref{1Nring}) is the Neveu-Schwarz ring. This gives rise to a sheaf, denoted by $\mathcal{A}_N$ for
some positive integer $N$. The only condition on the ring associated to each open
set, is that it be `holomorphic in $z$'. This is, in fact, quite a
subtle analytic condition, a discussion of which will be postponed for
a few paragraphs. Derivations
are now replaced by the sheaf of graded derivations, $\mathcal{D}^1\mathcal{A}_N$ 
which is a left $\mathcal{A}_N$ module. Accordingly, there is the
sheaf of graded one-forms, $\mathcal{D}_1\mathcal{A}_N$, which are $\mathcal{A}_N$ linear maps of
$\mathcal{D}^1\mathcal{A}_N$ into
$\mathcal{A}_N$. $\mathcal{D}_1\mathcal{A}_N$ is a right
$\mathcal{A}_N$ module. For full details of this construction of
derivations and one-forms, see \cite{kos}.

There is now a question of a preserved one-form. The basis of
differentials is now given by $(dz, d\theta_i)$. Rather than the one-form coming for free, it must now be
defined. This one-form will define a generalised conformal structure \cite{hodg}. Define the one form \cite{cohn} $\omega = dz - \sum_id\theta_i
\theta_i$. A transformation $(z, \theta_i) \mapsto (z', \theta_i')$ is
superconformal iff it is invertible and $\omega' = \omega \kappa(z, \theta_i)$ for $\kappa
\in \mathcal{A}_N$. All homomorphisms between open sets with
intersection are demanded to be superconformal. An alternative basis for
$\mathcal{D}_1\mathcal{A}_N$ is $(\omega, d\theta_i)$, which gives a
corresponding dual basis of $\mathcal{D}^1\mathcal{A}_N$, namely
$(\partial, D_i)$. Here, $\partial = \frac{\partial}{\partial z}$,
and $D_i = \frac{\partial}{\partial\theta_i} + \theta_i
\frac{\partial}{\partial z}$. This is a convenient basis to
work with. It can be readily shown from the superconformal condition,
that 
\begin{eqnarray}\label{1scfcndn}
&D_iz' - \sum_j \theta'_jD_i\theta'_j=0\\
&\kappa = \partial z' + \sum_j\theta_j'\partial\theta_j'
\end{eqnarray}
From here on summations are dropped, and summation convention should
be assumed. In particular,
considering infinitesimal transformations, once the $z'$
transformation is known, all the $\theta_i'$ transformations can
almost be
deduced. In the $N=1$ case, there is a $\mathbb{Z}_2$ ambiguity. For
higher $N$ this ambiguity becomes a continuous group, often interpreted
as a gauge group in the physics literature \cite{adem2}. $\kappa$ can also be given by a slightly different expression. The relation $\lbrack D_i, D_k \rbrack =
2\delta_{ik}\partial$, where the commutator is graded, and (\ref{1scfcndn}), are useful in
showing that $\det(D_i\theta_j')^2 = (\partial
z'+\theta_i'\partial\theta_i')^N = \kappa^N$. This is an expression in a
Grassmann ring over $\mathbb{C}$, so it is not obvious that one can divide or take
$N$-th roots. If the co-ordinate transformation on the intersection of
two open neighbourhoods $(z,\theta_i) \mapsto
(z',\theta_i')$ involves a scale factor $\kappa$, the inverse
transformation will induce a scale factor $\kappa'$, with
$\kappa\kappa' = \kappa'\kappa = 1$. Thus, $\kappa$ has a
unique \cite{ft1} inverse,
and both $\kappa$ and $\kappa'$ have a component that is pure complex
number, i.e. they have a non-zero `body'. Since an extended graded manifold
framework is being used, this is not as trivial a fact is it might
seem. $N$-th roots of $\kappa$ can now be
defined, as the binomial expansion around the `body', $\epsilon(\kappa)$. This expansion
is finite due to the nilpotency of $\kappa -
\epsilon(\kappa)$. One
finds that $\omega' = \omega (\det(D_i\theta_j'))^{\frac{2}{N}}\zeta^2$,
where $\zeta$ is an $N$-th root of
unity. Calculating the corresponding transformation on the basis of 
$\mathcal{D}^1\mathcal{A}_N$, one finds that $D_i' =
(D_i'\theta_j)D_j$, and that $(D_i\theta_j')^{-1} = (D_i'\theta_j)$.

There is a subtle point about the superconformal transformation. The
map 
\begin{eqnarray}\label{1spinxmfn}
z' = z, \quad\theta_i' = -\theta_i
\end{eqnarray}
is superconformal. If one
restricts to transformations $z' = z'(z)$ and uses the superconformal
condition to deduce how the $\theta_i$ can transform, the choice of
possible minus manifests itself as a choice of spin
structure. Considering the transformation (\ref{1spinxmfn}), one can
ask what functions of (\ref{1Nring}) are invariant under it. A
basis can be chosen for this ring, namely $\{ z^n, z^n\theta_i, \ldots,
z^n\theta_1\theta_2\dots\theta_N\}$, for $n\in\mathbb{Z}$. Without the minus sign, all these
basis elements are transformed to themselves. With the minus sign, one
finds that those elements with an odd number of $\theta_i$ obtain a
minus sign. Thus, only a subring is invariant. One can enlarge the
invariant subring by introducing square roots and choosing the minus
sign in the square root whenever one has an odd number of $\theta_i$
in a basis element. Consider now a subring of
\begin{eqnarray}\label{1N2ring}
B_{L'} \lbrack z^\frac{1}{2}, z^{-\frac{1}{2}}, \theta_1, \ldots, \theta_N \rbrack
\end{eqnarray}
which has as a basis $\{ z^n, z^{n+\frac{1}{2}}\theta_i,
z^n\theta_i\theta_j, z^{n+\frac{1}{2}}\theta_i\theta_j\theta_k
\ldots\}\quad n\in\mathbb{Z}$ and choose the negative square root
under a superconformal transformation. Now all the basis elements are
mapped to themselves under a superconformal transformation. This is
the Ramond ring. It should be noted that with the analytic definition
of functions that are `holomorphic in $z$' that will be used here (see below), the
Ramond ring introduces a branch cut. It should also be noted that this
construction can easily be extended to the case when some of the
$\theta_i$ have a minus sign in the transformation, and some do
not.

Already the need for a more analytic idea of what is going on is
apparent. This will be particularly important when the question of
contour integration arises. Full details of this approach can be found
in \cite{rog3}, \cite{rog1}. Many the details are not
mentioned here. Consider now a Grassmann algebra generated by $L$
generators, $B_L$. This algebra can be given a Banach algebra
structure, with the resulting topology being Hausdorff. This splits into an even and odd part, $B_L =
B_{L0}\oplus B_{L1}$ as a vector space over $\mathbb{C}$. Now the
$\theta$ `co-ordinates' take values in $B_{L1}$ and $z$ in
$B_{L0}$. In particular, $z$ can now have a nilpotent even part $s(z)$ (the
soul), but must have a proper complex number part $\epsilon(z)$ (the body). Using
the fact that a body exists, a `superdifferentiable'\cite{rog1}
function can be constructed. For $L' = \{$the smallest
integer not smaller than $\frac{L}{2} \}$, a continuation can be
specified, namely a continuation $Z_{L',L} : C^\infty(\epsilon(U),
B_{L'})\rightarrow G^\infty(U, B_L)$ for $U$ open in $B_L$. The
algebra $B_{L'}$ can be associated to the algebra $B_L$ by the
inclusion map $\iota_{L',L}:B_{L'}\rightarrow B_{L}$ which is the
algebra homomorphism that maps the
generators $\beta_i$ to $L'$ of the generators in $B_L$, and the unit
in $B_{L'}$ to the unit in $B_L$. Note,
$f(z)$ may be even or odd.
\begin{eqnarray}\label{zcontin}
Z_{L',L}(f)(z) =
\sum_{i=0}^L\frac{1}{i!}\iota_{L',L}\Big(f^{(i)}\big(\epsilon(z)\big)\Big)\times s(z)^i
\end{eqnarray}
Where $f^{(i)}$ denotes the i-th derivative. Now consider functions of
the variables $(z, \theta_i)$ that can be written as
\begin{eqnarray}
F(z, \theta_i) &=& Z_{L',L}(f_0)(z) + \sum_{i=1}^N
Z_{L',L}(f_i)(z)\theta_i + \ldots \nonumber\\
&=& \sum_{\mu = 0}^{2^L-1}Z_{L',L}(f_\mu)(z)\theta_\mu
\end{eqnarray} 
The statement that $F(z, \theta_i)$ is holomorphic in $z$ is the
statement all the $f_\mu(w)$, are holomorphic in the complex
variable $w$. These are examples of $GH^\infty$ functions. For a sheaf
to be $GH^\infty$, its restriction maps, $\rho_{UV}$, must also be
$GH^\infty$.

An interesting sheaf to look at is that formed by the one-form
$\omega$. It shares similar properties to the one-form $dz$ on an
ordinary Riemann sphere. Call the sections of this line bundle on the Riemann sphere
$\mathcal{O}_0(1)$. If the change of co-ordinates on an intersection
from one open set to another is $z\mapsto f(z)$, then the transition
function for an element of $\mathcal{O}_0(1)$ is $f'(z)$. In conformal
field theory, a primary field of weight $h$ can be thought of as a section of $\mathcal{O}_0(h) =
\mathcal{O}_0(1)^{\otimes h}$, where $h$ is a positive integer. In the
super case, this gives something similar. It should be pointed out
that $\omega$ will not give a line bundle. This would require the
typical fibre to be a free $B_{L'}$ module of rank one, rather than a
line. As a result, it must be regarded as a sheaf. This gives rise to
a sheaf of sections, $\mathcal{O}_N(1)$. On an intersection, the
function $(\det D_i\theta_j')^{\frac{2}{N}}\zeta^2$ is the sheaf's
homomorphism. This should be compared to the transition functions on a
line bundle. `Uncharged' primary
superfields \cite{dor1} can then be thought of as sections of
$\mathcal{O}_N(h)=\mathcal{O}_N(1)^{\otimes h}$. It should be noted that now tensor products are
taken over a graded ring, $\mathcal{A}_N$, so care must be taken with
signs in the tensor product. For example, consider a tensor product
between two graded left-$\mathcal{A}_N$ modules. Then
\begin{eqnarray}
f_1\otimes p f_2 = (-1)^{pf_1} pf_1\otimes f_2 \nonumber
\end{eqnarray}
where $p \in \mathcal{A}_N$, and exponents of $(-1)$ give the parity of
the associated element. Similar formulae should be used if one of the
two modules, or both are right-$\mathcal{A}_N$ modules, with the
obvious modifications in the exponent of $(-1)$.

There are other interesting sheaves also present. If one accepts the Berezin prescription
for integration, then the integration `measure' on the graded Riemann
sphere is $\omega \otimes_i D_i$, which has a transformation rule
$(\det D_i\theta_j')^{(\frac{2}{N}-1)}\zeta^2$. As was first noticed in
\cite{ks}, there is also a $O(N)$ group present. This can be regarded
as a sheaf in the following way. Since the $\{ D_i \}$, $i=1\ldots N$,
transform amongst themselves, one can consider the sheaf
of supercovariant derivatives. Then the transformation law on
intersections of open sets is $D_i' = (D_i' \theta_j)D_j$. Since the
superderivatives transform into one another, one can consider the sheaf
of supercovariant derivatives. Call the sections of this sheaf
$\mathcal{C}_N$. The matrices $D_i\theta_j'$ enjoy the following property
\begin{eqnarray}
(D_i \theta_j')(D_k \theta_j') = \delta_{ik}(\partial z' + 
\theta_j'\partial\theta_j') = \delta_{ik}\kappa = \delta_{ik}(\det D_i\theta_j')^\frac{2}{N}\zeta^2
\end{eqnarray}
The right hand side can be though of as an $N\times N$ identity matrix
multiplying the scale factor of the superconformal
transformation. Consider now a sheaf whose sections are
$\mathcal{O}(-\frac{1}{2})\otimes
\mathcal{C}_N$. Constructing $\mathcal{O}(-\frac{1}{2})$ requires
taking a square root, and is very analogous to taking the square root
of a line bundle. Hence, the choice of sign can be thought of as
choosing a spin structure. The group homomorphisms on
an intersection of open sets in this sheaf are given by
\begin{eqnarray}
M_{ij} = \frac{(D_i\theta_j')}{\sqrt{\kappa}}
\end{eqnarray}
and the sheaf itself is an
$\mathcal{A}_N$-module of rank $N$. Recall that on intersections,
$\kappa$ has an inverse $\kappa '$, and so $\kappa^{-\frac{1}{2}}$ is
well defined on this intersection. There is still a question of a
sign. Keeping in mind that $\theta_j' = \pm \theta_j$ is a
superconformal transformation, it is really the $\theta_j$ that one
would want to account for different spin structures, rather than
$\mathcal{O}(-\frac{1}{2})$. Therefore, it seems reasonable to choose
a plus sign for $\kappa^{-\frac{1}{2}}$. Regarding the new
homomorphism $M_{ij}$ as a matrix acting on a free module of rank $N$, it can be thought of as an element of
$O(N)$, with the entries being even Grassmann elements. The new sheaf then gives rise to a
fundamental representation of $O(N)$, and the $M_{ij}$ the
coefficients of a matrix with basis $E^{ij}$. Call this sheaf
$\mathcal{G}_N$. This can give rise to other sheaves which are also
$O(N)$ representations. 

The manner in which this is done parallels
what is often done for vector bundles, in particular frame
bundles and spin bundles. The reason this treatment can be applied is that $\mathcal{G}_N$ almost looks
like a vector bundle, the only hindrance being that the `typical
fibre' would be a $B_{L'}$ module rather than a vector space. Rather
than an abelian group of rank $N$ being associated to each open set,
one can instead associate a group element of $O(N)$, just as is done
with frame bundles with principal bundles, and retain the same $M_{ij}$. This gives rise to a
sheaf $\widetilde{\mathcal{G}}_N$. Considering, now a different
representation $\rho$ of $O(N)$ gives rise to a sheaf homomorphism
(albeit of non-abelian groups)
\begin{eqnarray}\label{nonabhomo}
\sigma : (\mathbb{P}_1, \widetilde{\mathcal{G}}_N )\rightarrow
(\mathbb{P}_1, \widetilde{\mathcal{G}}_N^\rho )
\end{eqnarray}
where the group homomorphisms are now given by $M_{ij}\rho
(E^{ij})$. Since the representation $\rho$ has a vector space $V$
associated to it, one can consider the sheaf which has the group
homomorphisms given by $M_{ij}\rho (E^{ij})$, and stalk $V$, and denote
the sections of this sheaf by $\mathcal{R}(G_N)$. An $O(N)$-extended
primary superfield, first introduced in \cite{ks}, can then be defined
as a section of $\mathcal{O}_N(h)\otimes\mathcal{R}(G_N)$.

\section{Contour Integration}
Since one wishes to do conformal field theory in the setting presented
above, a sensible question to ask is what closed contour integrals
will look like, given the set of analytic functions
(\ref{zcontin}). All functions on a given open set look like
\begin{eqnarray}
f(z) = Z_{L',L}(f_0)(z) = \sum_{i=0}^L
\frac{1}{i!}\iota_{L',L}(f_0^{(i)}(\epsilon (z)))s(z)^i 
\end{eqnarray}
where $f_0 \in C^\infty (\mathcal{U}, B_{L'})$. This should be compared to
the usual notion of a Taylor expansion, around $\epsilon(z)$. Note that if $L'=0$,
$B_{L'}=\mathbb{C}$, and $H^\infty$ functions are retrieved. In
the following, the $\iota_{L',L}$ will be suppressed(for clarity). It is a linear
map, so one can see that the following workings are unaffected. The
contour integral $\oint_{C_z}f(z)dz$ needs to be considered. By the
definition above, if $z$ is an ordinary complex number (i.e. has no soul), one finds
\begin{eqnarray}
f(z) = \sum_{i=0}^L \frac{1}{i!} f_0^{(i)}(\epsilon(z))s(z)^i = f_0(z)
\end{eqnarray}
If the even co-ordinate, $w$, were to have soul as well as body, it would give
an element of an even Grassmann algebra over a complex field, when
evaluated at a point. Hence one can consider $w$ itself as being
parameterised by a complex number $z$. Now consider a parameterisation $w =
g(z)=b(z)+u(z)$, where $\epsilon(w) = b(z)$, $s(w) = u(z)$.   
\begin{eqnarray}
f(w) &=& \sum_{i=0}^L \frac{1}{i!}f_0^{(i)}(\epsilon(w))s(w)^i
\nonumber\\
&=& \sum_{i=0}^p \frac{1}{i!}(f_0^{(i)} \circ b)(z)u(z)^i \nonumber \\
&=& (f \circ g)(z)
\end{eqnarray}
where $p<L$ is the integer such that $u(z)^p \neq 0$, $u(z)^{p+1} =
0$. Using the definition of a contour integral given in \cite{rog2},
with $C_w = g(C_z)$,
\begin{eqnarray}
\oint_{C_w} f(w)dw &=& \oint_{C_z}(f\circ g)(z)g'(z)dz \nonumber\\
&=& \oint_{C_z}\Big( \sum_{i=0}^p \frac{1}{i!}(f_0^{(i)} \circ
b)(z)u(z)^i \Big) \Big(\frac{d}{dz}b(z) + \frac{d}{dz}u(z)\Big) dz \nonumber\\
&=& \oint_{C_z}\Big( \frac{d}{dz}b(z)\Big)(f_0\circ b)(z) +
\sum_{i=1}^p\frac{1}{i!}\Big( \frac{d}{dz}b(z)\Big)(f^{(i)}\circ
b)(z)u(z)^i\nonumber\\
&& \qquad + \sum_{i=0}^p\frac{1}{i!}(f^{(i)}\circ b)(z)u(z)^i\Big(
\frac{d}{dz}u(z)\Big) dz
\end{eqnarray}
All that has be done above is put all the definitions in and split up
some summations. Note in the first summation, the chain rule can be
used on the function $b(z)$, and in the second summation, the chain rule can be used on $u(z)$, giving
\begin{eqnarray}
\oint_{C_w} f(w)dw &=& \oint_{C_z} \Big(
\frac{d}{dz}b(z)\Big)(f_0\circ b)(z) +
\sum_{i=1}^p\frac{1}{i!}\Big(\frac{d}{dz}(f^{(i-1)}\circ
b)(z)\Big)u(z)^i\nonumber\\
&&\qquad + \sum_{i=0}^p\frac{1}{(i+1)!}(f^{(i)}\circ
b)(z)\frac{d}{dz}\Big(u(z)^{i+1}\Big)\quad dz\nonumber\\
&=&  \oint_{C_z} \Big(\frac{d}{dz}b(z)\Big)(f_0\circ b)(z) +
\sum_{i=1}^p\frac{1}{i!}\frac{d}{dz}\Big( (f^{(i-1)}\circ b)(z)\cdot
u(z)^i \Big)\nonumber\\
&&\qquad + \frac{1}{(p+1)!}(f^{(p)}\circ b)(z) \frac{d}{dz}\Big(u(z)^{p+1}\Big)\quad dz
\end{eqnarray}
The last term is in fact zero, due to the nilpotency of $u(z)$. The term under the
summation is a total derivative. As such,
integrated around a closed contour, it vanishes identically. All that remains is
\begin{eqnarray}
\oint_{C_w} f(w)dw &=& \oint_{C_z}(f_0 \circ b)(z)b'(z)dz
\end{eqnarray}
Thus, the contour
integral can formally be treated as an integral in a normal complex
number. It should be noted that all that has been used in
this calculation is the chain rule and product rule over $C^\infty$ functions.

\section{The Preserved One-Form and Ramond Fields}
Whilst generalising the bosonic setting in the previous sections, it
was found that rather than a preserved one-form coming for free, it
had to be specified. One could ask, what happens if another one-form
is specified. In \cite{kac3}, other one-forms were considered. It was found that if
one wanted a $\mathbb{Z}$-graded algebra, a one-form of the form $dz -
d\theta_i f(z)\theta_i$ had to have $f(z)=z^n$. By making a change of
variables, \cite{kac3} then shows that one only need consider the
cases $n=0,1$.

Consider, now, a different preserved one-form, namely
$\omega = dz - d\theta_i z\theta_i$. The dual derivations
to $(\omega, \theta_i)$ are $(\partial, D_i)$, where $D_i =
\frac{\partial}{\partial\theta_i} + z\theta_i\frac{\partial}{\partial
z}$. Now one finds that $\lbrack D_i, D_j \rbrack =
2\delta_{ij}z\partial$. Requiring that under a transformation,
$\omega' = \omega\kappa(z,\theta_i)$ yields
\begin{eqnarray}
&D_j z' - z'\theta_i'D_j\theta_i' = 0\\
&\kappa = \partial z' + z'\theta_i'\partial\theta_i' =
(\frac{z'}{z})(\det D_i\theta_j')^\frac{2}{N}\zeta^2\\
&D_i' = (D_i'\theta_j)D_j \\
&(D_i\theta_j')(D_k\theta_j') = \delta_{ik}(\det D_i\theta_j')^{\frac{2}{N}}\zeta^2
\end{eqnarray}
One can ask what an element of $\mathcal{O}_N(h)$ may transform like,
and what is the algebra of infinitesimal transformations associated
to it. Consider the case $N=1$. The field has a transformation rule
under $(z,\theta)\mapsto g(z,\theta) = (z',\theta')$
\begin{eqnarray}\label{altram}
(U_g^{-1}\Phi U_g)(z',\theta') =  \Phi'(z', \theta') = \Phi(z, \theta)\Big( \Big(\frac{z'}{z}\Big)
(D\theta')^2\Big) ^{-h}
\end{eqnarray}
The superconformal condition, namely preserving the new one-form $w$, imposes two types of transformation, a
bosonic and a fermionic one. On the co-ordinates, $(z, \theta)$,
the infinitesimal transformations are given by 
\begin{eqnarray}
(z, \theta)&\mapsto &(z+az^{n+1}, \theta +
  a\frac{n}{2}z^n\theta)\qquad\textrm{and}\nonumber\\
(z, \theta)&\mapsto &(z+\epsilon\theta z^{r+1}, \theta - \epsilon z^r)
\end{eqnarray}
where $r,n \in \mathbb{Z}$. These provide a basis for all
infinitesimal transformations. Each one induces a transformation on
the field, (\ref{altram}), with $a(n) = az^{n+1}$, $\epsilon
(r)=\epsilon\theta z^{r+1}$
\begin{eqnarray}\label{inframondphi}
& \delta_{a(n)}\Phi (z,\theta) = -a\Big(z^{n+1}\partial_z +
\frac{n}{2}z^n\theta\partial_\theta +
h(n+1)z^n\Big)\Phi(z,\theta)\nonumber\\
& \delta_{\epsilon(r)}\Phi (z,\theta)= -\epsilon\Big( \theta z^{r+1}\partial_z -
z^r\partial_\theta + h(2r+1)\theta z^r\Big)\Phi(z,\theta)
\end{eqnarray}
where $n,r \in \mathbb{Z}$. These differential operators give rise to commutation relations
\begin{eqnarray}\label{inframondalg}
&\lbrack \delta_{a_1(m)}, \delta_{a_2(n)} \rbrack =
(m-n)\delta_{a_2a_1(m+n)}\nonumber\\
&\lbrack \delta_{\epsilon_1(r)}, \delta_{\epsilon_2(s)} \rbrack =
2\delta_{\epsilon_2\epsilon_1(r+s)}\nonumber\\
&\lbrack \delta_{a(m)}, \delta_{\epsilon(r)} \rbrack = (\frac{m}{2} -
r)\delta_{\epsilon a(m+r)}
\end{eqnarray}
This gives a representation of the Ramond algebra. Note that no branch
cut has been introduced. Another thing to note is that $\delta_{a(0)}$
gives a $l_0$ operator, which says that $\theta$ scales like a field
of weight zero, rather than a field of weight half. As a result, the
expansion of $\Phi(z,\theta)$ is now taken to be 
\begin{eqnarray}
\Phi(z,\theta) = \phi_0(z) + \theta\phi_1(z) =
\sum_{m\in\mathbb{Z}}\phi_{0m}z^{-m-h} +
\theta\sum_{m\in\mathbb{Z}}\phi_{1m}z^{-m-h}\nonumber
\end{eqnarray}
Using this expansion, and writing the transformation of a field as
\begin{eqnarray}\label{altram2}
(U_g\Phi U_g^{-1})(z,\theta) = \Phi(z', \theta')\Big( \Big(\frac{z'}{z}\Big)
(D\theta')^2\Big) ^{h}\qquad
 U_g = \exp(a_nL_n + \epsilon_rG_r)
\end{eqnarray}
one can find the action of the algebra on the modes of $\Phi$
as\cite{ft2} 
\begin{eqnarray}
&\lbrack L_n, \phi_{0m} \rbrack = \big(
(h-1)n-m\big)\phi_{0m+n}\qquad
\lbrack L_n, \phi_{1m} \rbrack = \big(
(h-\frac{1}{2})n-m\big)\phi_{1m+n}\nonumber\\ \nonumber\\
&\lbrack G_r, \phi_{0m} \rbrack = \phi_{1m+r}\qquad
\lbrack G_r, \phi_{1m} \rbrack = \big( (2h-1)r-m\big)\phi_{0m+r}\nonumber
\end{eqnarray}
These are precisely the commutation relations one obtains from the
$N=1$ Ramond OPEs from the usual method of introducing a branch
cut\cite{lust}. Rewriting the commutation relations
(\ref{inframondalg}) in a more familiar way, and inserting the unique
central extension \cite{kac3}, the algebra can be written down
\begin{eqnarray}
&\lbrack L_m, L_n \rbrack = (m-n)L_{m+n} +
\frac{C}{6}m(m^2-1)\delta_{m+n,0}\nonumber\\ \nonumber\\
&\lbrack L_m, G_r \rbrack = (\frac{m}{2}-r)G_{m+r}\qquad \lbrack G_r,
G_s \rbrack = 2L_{r+s} + \frac{2C}{3}(r^2-\frac{1}{4})\delta_{r+s,0}
\end{eqnarray}
Influenced by the form of the infinitesimal changes
(\ref{inframondphi}), OPEs can be postulated that give the above
commutation relations, which read as
\begin{eqnarray}
L(w)\phi_0(z) &\sim & \Big(\frac{\partial}{(w-z)} +
\frac{h}{(w-z)^2}\Big) \phi_0(z) \nonumber\\
L(w)\phi_1(z) &\sim & \Big(\frac{\partial}{(w-z)} +
\frac{h+\frac{1}{2}}{(w-z)^2} - \frac{1}{2z(w-z)}\Big) \phi_0(z) \nonumber\\
G(w)\phi_0(z) &\sim & \frac{1}{z(w-z)}\phi_1(z)\nonumber\\
G(w)\phi_1(z) &\sim & \Big(\frac{\partial}{(w-z)} + \frac{2h}{(w-z)^2} -
\frac{h}{z(w-z)}\Big) \phi_0(z)\\
L(w)L(z) &\sim & \frac{C}{(w-z)^4} + \frac{2L(z)}{(w-z)^2} +
\frac{\partial L(z)}{(w-z)}\nonumber\\
L(w)G(z) &\sim & \frac{\partial G(z)}{(w-z)} + \frac{3G(z)}{2(w-z)^2} -
\frac{G(z)}{2z(w-z)}\nonumber\\
G(w)G(z) &\sim & \frac{2zL(z)}{(w-z)} + \frac{2C}{3}\Big(
\frac{2z}{(w-z)^3} + \frac{1}{(w-z)^2} - \frac{1}{4z(w-z)}\Big)
\end{eqnarray}
where $L(z) = \sum_n L_nz^{-n-2}$, $G(z) = \sum_r G_rz^{-r-1}$,
$z,r\in\mathbb{Z}$. This final set of OPEs demonstrate the drawbacks
of the more abstract construction used in this paper of a Conformal
Field Theory (namely via a section of a sheaf over some manifold),
compared to the more
usual approach of a free field realization. One can calculate the
infinitesimal transformations of the field (the section obtained), and
show the transformations close as a lie algebra. One then has to `work
backwards' and try and construct OPEs and central charge terms that
agree with the transformations and lie algebras calculated. It would
be interesting to see if the Ramond field (\ref{altram}) could be
realized via a free field realization where
usually one finds central terms and OPEs are explicitly calculable.

\section{Classical $N=3$ Algebra}
Consider now that case of preserving the usual one-form,
$dz-d\theta_i\theta_i$, with three Grassmann variables. The
superconformal condition is then (\ref{1scfcndn}). Using a notation of $Z =
(z,\theta_i)$, and writing a superconformal transformation as $Z
\mapsto  g(Z)$, a representation of the group can be constructed
via $\mathcal{A}_3$, namely $U_gf(Z) = (f\circ g^{-1})(Z)$. The infinitesimal
transformations can be calculated, and a Lie superalgebra
constructed. The infinitesimal transformations take the form of vector
fields acting on functions.

The most general infinitesimal transformation on the $z$ co-ordinate
is 
\begin{eqnarray}
z \mapsto z + af(z, \theta_1, \theta_2, \theta_3) + \epsilon h(z,
\theta_1, \theta_2, \theta_3)\nonumber
\end{eqnarray}
for $f$ ($h$) some even (odd)
function, and $a$ ($\epsilon$) infinitesimal and of even (odd) parity. The functions
$f$ have analogues for transformations in the
$\theta_i$ co-ordinates. Breaking up $f$ into superfield components
gives eight different types of transformation. 
\begin{eqnarray}\label{infzxmfn}
z \mapsto z + a(z) + \alpha_i(z)\theta_i +
\frac{1}{2}a_{ij}(z)\theta_i\theta_j + \alpha_{123}(z)\theta_1\theta_2\theta_3
\end{eqnarray}
The possible
transformations are forced into only these eight types, and not some
mix between them, by the superconformal condition. 

An infinitesimal transformation most generally reads
\begin{eqnarray}
z' = z+\delta z \qquad \theta_j' = \theta_j + \delta\theta_j
\end{eqnarray}
On substituting into (\ref{1scfcndn}), one finds that the
superconformal condition reads
\begin{eqnarray}
D_i \delta z = \delta\theta_i + \sum_{j=1}^3 \theta_jD_i\delta\theta_j
\end{eqnarray}

i.e three equations, with three unknowns once $\delta z$ has been
specified. A basis for the infinitesimal $z$ transformations is easily
found, which is $\delta z = \epsilon\theta_1\theta_2\theta_3
z^{n+\frac{1}{2}}$, $\delta z = a\theta_i\theta_j z^{n+1}$
for $i<j$, $\delta z = \epsilon\theta_i z^{n+\frac{1}{2}}$, and
$\delta z = az^{n+1}$. Given these eight types of transformation,
precisely what the corresponding $\delta\theta_i$ are, modulo possible
$\delta\theta_i$ if $\delta z = 0$, can be
calculated explicitly. The case when $\delta z =0$ is taken care of by
the $t_n^i$ generators below. Using this procedure, the infinitesimal
generators of the $N=3$ algebra can be calculated. The results
are quite hefty, but the actual transformations give an
intuitive idea of what each element of the algebra actually
does. Summation convention is used in the following.
\begin{eqnarray}\label{lnvf}
&z \mapsto z + az^{n+1} \quad \theta_i \mapsto \theta_i + a\frac{1}{2}(n+1)\theta_i
z^{n}\nonumber \\ & \Rightarrow l_m = -z^m \Big( z
\frac{\partial}{\partial z} +
\frac{1}{2}(m+1)\theta_i\frac{\partial}{\partial \theta_i} \Big)
\end{eqnarray}
gives a vector field corresponding to an infinitesimal transformation
when only $a(z)$ is non-zero in (\ref{infzxmfn}). There are then the
three single $\theta$ terms, which can be found by considering the
case when only one $\alpha_i(z)$ is non-zero.
\begin{eqnarray}
z &\mapsto & z - \epsilon \theta_1 z^{r+\frac{1}{2}} \nonumber\\ \theta_1
&\mapsto &\theta_1 + \epsilon z^{r+\frac{1}{2}} \nonumber \\
\theta_2 &\mapsto &\theta_2 - \epsilon (r+\frac{1}{2})\theta_1\theta_2
z^{r-\frac{1}{2}} \nonumber \\ \theta_3 &\mapsto &\theta_3 - \epsilon
(r+\frac{1}{2})\theta_1\theta_3 z^{r-\frac{1}{2}} \nonumber
\end{eqnarray}
gives rise to the vector field
\begin{eqnarray}
 g_r^1 =
z^{r-\frac{1}{2}}(z\theta_1\frac{\partial}{\partial z} -
z\frac{\partial}{\partial \theta_1} +
(r+\frac{1}{2})\theta_1\theta_2\frac{\partial}{\partial \theta_2} +
(r+\frac{1}{2}) \theta_1\theta_3\frac{\partial}{\partial \theta_3}) \nonumber
\end{eqnarray}
Similarly
\begin{eqnarray}
g_r^i = z^{r-\frac{1}{2}}(z\theta_i\frac{\partial}{\partial z} -
z\frac{\partial}{\partial \theta_i} +
(r+\frac{1}{2})\theta_i\theta_j\frac{\partial}{\partial \theta_j}) \nonumber
\end{eqnarray}
It is worth noting that if one were not working on an extended graded
manifold, but on a graded manifold (c.f. (\ref{1Nring}),
(\ref{oldring})), then one would not be able to obtain the above
vector field. The same statement holds for $\psi_r$ below.

There are three double $\theta$ terms, e.g. $\theta_1 \theta_2$ gives $t_n^3$
\begin{eqnarray}
z &\mapsto & z  \nonumber\\
\theta_1 &\mapsto & \theta_1 + a \theta_2 z^{n+1} \nonumber\\
\theta_2 &\mapsto & \theta_2 - a \theta_1 z^{n+1} \nonumber\\
\theta_3 &\mapsto & \theta_3 + a(n+1)\theta_1\theta_2\theta_3 z^n \nonumber
\end{eqnarray}
A similar calculation applies to $t_n^1$ and $t_n^2$
\begin{eqnarray}
t_m^i = z^{m-1}(z \epsilon_{ijk} \theta_j \frac{\partial}{\partial\theta_k} - m
\theta_1\theta_2\theta_3\frac{\partial}{\partial\theta_i}) \nonumber
\end{eqnarray}
These transformations leave the $z$ component unaltered, and as such
have sometimes been interpreted in the physics literature \cite{adem2} as a
gauge group.
The final term is similarly calculated, and is the three $\theta$
transformation
\begin{displaymath}
\psi_r = -
z^{r-\frac{1}{2}}(\theta_1\theta_2\theta_3\frac{\partial}{\partial z}
+ \frac{1}{2}\epsilon_{ijk}\theta_i\theta_j\frac{\partial}{\partial \theta_k})
\end{displaymath}
These vector fields, similarly calculated in \cite{kac2}, \cite{brem}
then give rise to the commutation relations for the $N=3$ algebra
without central extension.
\begin{eqnarray}\label{n3class}
& \lbrack t_m^i, t_n^j \rbrack = -\epsilon_{ijk}t_{m+n}^k \quad \lbrack t_m^i, \psi_s
\rbrack = 0 \quad \lbrack t_m^i, g_r^j \rbrack =
\delta_{ij}m\psi_{r+m} - \epsilon_{ijk}g_{r+m}^k\nonumber\\ \nonumber\\
& \lbrack l_m, \psi_s \rbrack = -(\frac{m}{2}+s)\psi_{m+s} \quad
\lbrack l_m, t_n^i \rbrack = -nt_{m+n}^i \quad \lbrack g_r^i, \psi_s
\rbrack = t_{r+s}^i \nonumber \\ \nonumber\\
&\lbrack l_m, g_r^i \rbrack = (\frac{m}{2}-r)g_{r+m}^i \quad \lbrack
g_r^i, g_s^j \rbrack = 2\delta_{ij} l_{r+s} +
\epsilon_{ijk}(r-s)t_{r+s}^k \nonumber\\ \nonumber\\ &  \lbrack l_m,
l_n \rbrack = (m-n)l_{m+n} \quad \lbrack \psi_m, \psi_n \rbrack = 0
\end{eqnarray} 
Note in particular that the $t_n^i$ form an $su(2)$ loop algebra,
which will be enhanced by a central extension in the quantum case to
give an affine $su(2)$ algebra. One implication of this is that the
representation theory will have to be very different to that of the $N=2$
case, where a $u(1)$ loop algebra appeared. The highest weight state
must also be an $su(2)$ highest weight state. Since $U(1)$ is abelian,
all its irreducible representations are one dimensional. The upshot of
this is that the OPE can be easily adapted by including one more
quantum number. Since $SU(2)$ is non-abelian, it will be seen that $su(2)$ generators will
appear in the OPE.

\section{Quantum $N=3$ Algebra}
The quantum $N=3$ algebra was calculated from a Lagrangian approach,
and canonically quantised in \cite{adem2}. Whilst this section may
look very technical, it should be stressed that essentially the same
procedure is being used as in the well documented bosonic case, where the
starting point is a section of a sheaf, namely
$\mathcal{O}_0(h)$. One plays the same game, but now uses the section
$\mathcal{O}_N(h)\otimes\mathcal{R}(G_N)$.
Since it is defined covariantly, one can then
write down how it transforms. This then gives rise to infinitesimal
transformations $\delta\Phi$, which close as a lie algebra. These
relations can be written in terms of an operator $\mathbb{T}$ acting
on $\Phi$, giving an OPE. From the $\delta\Phi$, an
ansatz for the OPE of $\mathbb{T}$ with itself can be inferred. The
action of the quantum algebras on primary fields is inherent in the
$\mathbb{T}\Phi$ OPE. The commutation relations of the
quantum algebra are then inherent in the $\mathbb{T}\mathbb{T}$ OPE as the modes. What
must be checked from the first OPE, is that the primary superfield does
indeed yield a highest weight vector.

Recall that for a primary field in the bosonic case, one performs a
diffeomorphism from the Riemann sphere to itself that obeys the
conformal condition, and looks at how the primary field
transforms. More precisely, one considers, a diffeomorphism $f$
\begin{eqnarray}
f : &\mathbb{P}_1 \rightarrow &\mathbb{P}_1\nonumber\\
& z \mapsto & f(z) = z'
\end{eqnarray}
with the conformal condition
\begin{eqnarray}
(f^*dz) = dz\kappa (z)
\end{eqnarray}
One then calculates how $\mbox{\boldmath $\phi$}\in\mathcal{O}(h)$
transforms under a pull-back, where $\mbox{\boldmath $\phi$}$ in local co-ordinates is
$\phi(z)dz^{\otimes h}$
\begin{eqnarray}
(f^*\mbox{\boldmath $\phi$})(z) = \kappa^{h}(\phi\circ f)(z)
  dz^{\otimes h} = \Big(\frac{dz'}{dz}\Big)^{h}\phi(z') dz^{\otimes h}
  =: \phi'(z) dz^{\otimes h}
\end{eqnarray}
yielding the transformation law
\begin{eqnarray}
(U_g\phi U_g^{-1})(z)= \Big(\frac{dz'}{dz}\Big)^{h}\phi(z') = \phi'(z)
\end{eqnarray}
For the graded case, one has to consider an invertible sheaf morphism
\begin{eqnarray}
f : &(\mathbb{P}_1, \mathcal{A}_N) \rightarrow &(\mathbb{P}_1, \mathcal{A}_N)\nonumber\\
& Z=(z, \theta_i) \mapsto & Z'=(z', \theta_i')
\end{eqnarray}
such that $f$ (as well as $f^{-1}$) has a $GH^\infty$ action on the functions $\mathcal{A}_N$, and
obeys the conformal condition
\begin{eqnarray}
(f^*\omega) = \omega\kappa (Z)
\end{eqnarray}
The transformation rule for the components of $\mbox{\boldmath
  $\Phi$}\in\mathcal{O}_N(h)\otimes\mathcal{R}(G_N)$ under a pull-back
  are then given by
\begin{eqnarray}
\Phi '(Z) =
  \kappa^{h}\frac{(D_i\theta_j')}{\sqrt{\kappa}}g^{ij}(\Phi\circ f)(Z)
\end{eqnarray}
The $g^{ij}$ are, up to a discrete
subgroup, a representation of the lie group
$O(N)$. The $g^{ij}$ explicitly realize the map (\ref{nonabhomo}). This formula matches that found in \cite{ks} for how a primary
superfield transforms. One now writes down the transformation law as
\begin{eqnarray}
(U_g\Phi U_g^{-1})(Z) =
  \kappa^{h}\frac{(D_i\theta_j')}{\sqrt{\kappa}}g^{ij}\Phi(Z')
\end{eqnarray}
and parameterise the group action infinitesimally by
\begin{eqnarray}
U_g = \exp (a_nL_n + \alpha_r^i G_r^i + b_n^i T_n^i +
\beta_r\psi_r)\nonumber\\
Z' = \exp (a_n l_n + \alpha_r^i g_r^i + b_n^i t_n^i + \beta_r\psi_r)Z\nonumber
\end{eqnarray}
in a completely analogous way to (\ref{altram2}), to obtain the
commutators of the superVirasoro operators on a primary field. For the
$N=3$ case this yields (\ref{n3quant}). One must now work backward to try and
construct an OPE between a stress-energy tensor and primary superfield
that yield these commutators.

For the $N=3$ case, the stress-energy tensor will be weight
$\frac{1}{2}$, and have superfield
decomposition\cite{ft3}
\begin{eqnarray}\label{n3se}
\mathbb{T}(Z) = \theta_1\theta_2\theta_3L(z) +
\frac{1}{2}\epsilon_{ijk}\theta_i\theta_j G^k(z) + \theta_iT^i(z)
+\psi(z)
\end{eqnarray}
An OPE for $N=3$ is found in\cite{ck}, that, on
contour integration, gives rise to the infinitesimal transformations
of a primary superfield \cite{ks}. With $Z_1 = (w,\chi_i)$,
$Z_2=(z,\theta_i)$, this reads
\begin{eqnarray}\label{n3ope}
\mathbb{T}(Z_1)\Phi(Z_2) &\sim &
\frac{h\theta_{12,1}\theta_{12,2}\theta_{12,3}}{Z_{12}^2}\Phi(Z_2) + 
\frac{\theta_{12,1}\theta_{12,2}\theta_{12,3}}{Z_{12}}\partial_w\Phi(Z_2)
+
\nonumber\\&&\frac{\epsilon_{ijk}\theta_{12,i}\theta_{12,j}D_{2,k}}{4Z_{12}}\Phi(Z_2)
+ \frac{\theta_{12,i}J_i}{Z_{12}}\Phi(Z_2)
\end{eqnarray}
where
\begin{eqnarray}
D_{2,i} = \frac{\partial}{\partial \theta_i} +
\theta_i\frac{\partial}{\partial z}\qquad Z_{12}=(w-z-\chi_i\theta_i)\qquad \theta_{12,i}=(\chi_i-\theta_i)\nonumber
\end{eqnarray}
Where the $J_i$ form an $su(2)$ algebra\cite{ft4}. The field,
$\Phi(Z)$, now also lives in an $su(2)$ representation, say $\mathcal{V}$. It is in fact
an $su(2)$ highest weight. $\mathbb{T}$ can then be thought of as
being an endomorphism of $\mathcal{V}$, e.g. explicitly with $su(2)$
indices $\mathbb{T}(Z_1)^a_{~b}\Phi(Z_2)^b$. This OPE is effectively a non-abelian version of the $q$ term
appearing in the $N=2$ case. In its place, another quantum number
appears, which is the $J_3$ eigenvalue. On the representation space, the action of
$T_0^i$ on a highest weight state is identified with that of
$J_i$. The OPE can be split up into $\theta$
components, according to (\ref{n3se}), and modes be taken of each
of the operators, $L(z), G^i(z), T^i(z), \psi(z)$, giving the
formulae (\ref{n3quant}), as required. Note in particular, how the classical algebra
appears in the relations again. The extra terms are the $h$ terms,
which will give the $L_0$ eigenvalue $h$. The other extra terms, the
$J_i$, will give an action of $su(2)$ on the primary field, and hence
on the highest weight, which we
know must be required from the classical analysis (\ref{n3class}),
where a $su(2)$ loop algebra appeared. 
\begin{eqnarray}\label{n3quant}
\lbrack L_m, \Phi(Z) \rbrack &=& z^m \Big( h(m+1) + z\partial_z +
\frac{1}{2}(m+1)\theta_i\partial_{\theta_i} + 
\frac{1}{2z}m(m+1)\epsilon_{ijk}\theta_i\theta_j J_k \Big) \Phi(Z)
\nonumber\\ \nonumber\\
\lbrack G_s^i, \Phi(Z) \rbrack &=& -z^{(s-\frac{1}{2})}\Big(
h(s+\frac{1}{2})\theta_i + \frac{1}{2}\theta_i z\partial_z -
\frac{1}{2}z\partial_{\theta_i} +
\frac{1}{2}(s+\frac{1}{2})\theta_i\theta_j\partial_{\theta_j} + \nonumber\\&&\qquad
(s+\frac{1}{2})\big( \epsilon_{ijk}\theta_j J_k \big) - 
\frac{1}{z}(s^2-\frac{1}{4})\theta_1\theta_2\theta_3 J_i \Big)\Phi(Z)
\nonumber\\ \nonumber\\
\lbrack T_m^i, \Phi(Z) \rbrack &=& z^{(m-1)}\Big( \frac{mh}{2}\epsilon_{ijk}\theta_j\theta_k -
\epsilon_{ijk}\frac{1}{2}z\theta_j\partial_{\theta_k} +
\frac{1}{2}m\theta_1\theta_2\theta_3\partial_{\theta_i} + zJ_i - \nonumber\\&&\qquad
m\big(\theta_i\theta_k J_k\big) \Big) \Phi(Z)
\nonumber\\ \nonumber\\
\lbrack \psi_s, \Phi(Z) \rbrack &=& z^{(s-\frac{1}{2})}\Big(
-\frac{h}{z}(s-\frac{1}{2})\theta_1\theta_2\theta_3 +
\frac{1}{2}\theta_1\theta_2\theta_3\partial_z +
\frac{1}{4}\epsilon_{ijk}(\theta_i\theta_j\partial_{\theta_k} - \theta_iJ_i\Big)\Phi(Z)
\end{eqnarray}
Note that
\begin{eqnarray}\label{svertextrans}
[L_{-1}, \Phi(Z)] = \partial_z \Phi(Z) \qquad [G_{-\frac{1}{2}}^i,
\Phi(Z)] = \frac{1}{2}(\partial_{\theta_i} - \theta_i
\partial_z) \Phi(Z) 
\end{eqnarray}
In particular, $L_{-1}$ acts as a translation in $z$ and
$G_{-\frac{1}{2}}^i$ as a super-translation in the respective
$\theta_i$ direction. This allows vertex operators to be used, and an
operator-state mapping employed \cite{kac}\cite{ft5}. In the bosonic theory, vertex
operators are characterised uniquely by their action on a vacuum
$|0\rangle$, which is annihilated by the raising operators, $\{ L_n:
n \geq -1 \}$. The vacuum cannot be invariant under the whole symmetry
algebra without implying vanishing of the central extension. This generalises to $N=3$, so that now
$|0\rangle$ is annihilated by $ \{L_n, G_r^i, T_m^i, \psi_s: n\geq -1
,r\geq -\frac{1}{2}, m \geq 0, s \geq \frac{1}{2} \}$. To get the state associated to any
vertex operator $\Phi(Z)$, one looks at
$\lim_{Z\to  0}\Phi(Z)|0\rangle$. Given this, it can be seen from the
relations (\ref{n3quant}) that the action of the raising operators on
$|\Phi\rangle$ is zero, e.g. for $\{ L_m: m>0 \}$
\begin{eqnarray}
\lim_{Z\to 0}[L_m, \Phi(Z)]|0\rangle = 0 \nonumber
\end{eqnarray}
The action of $L_0$ is given by
\begin{eqnarray}
&\lbrack L_0, \Phi(Z) \rbrack = \big(h + z\partial_z +
\frac{1}{2}(\theta_1\partial_{\theta_1} + \theta_2\partial_{\theta_2}
+ \theta_3\partial_{\theta_3})\big) \Phi(Z) \nonumber\\
&\Rightarrow \lim_{Z\to 0} \lbrack L_0, \Phi(Z) \rbrack |0\rangle =
h|\Phi\rangle  = L_0|\Phi\rangle
\end{eqnarray}
The action of $T_0^i$ is given by
\begin{eqnarray}
&\lbrack T_0^1, \Phi(Z) \rbrack = (\frac{1}{2}\theta_3
\partial_{\theta_2} - \frac{1}{2}\theta_2 \partial_{\theta_3} + J_1)
\Phi(Z) \nonumber\\
&\lbrack T_0^2, \Phi(Z) \rbrack = (\frac{1}{2}\theta_1
\partial_{\theta_3} - \frac{1}{2}\theta_3 \partial_{\theta_1} + J_2)
\Phi(Z) \nonumber\\
&\lbrack T_0^3, \Phi(Z) \rbrack = (\frac{1}{2}\theta_2
\partial_{\theta_1} - \frac{1}{2}\theta_1 \partial_{\theta_2} + J_3)
\Phi(Z) \nonumber\\
&\Rightarrow \lim_{Z\to 0} \lbrack T_0^i, \Phi(Z) \rbrack |0\rangle  =
\lim_{Z\to 0} (J_i\Phi )(Z) |0\rangle = J_i|\Phi\rangle
= T_0^i|\Phi\rangle
\end{eqnarray}
where the vacuum is $T_0^i$ invariant.

Hence, on the highest weight state, the
$T_0^i$ can be identified with the $J_i$, so that $T_0^3$
gives rise to the $q$ eigenvalue, and $J^+ = J_0^1+iJ_0^2$ annihilates
$|\Phi\rangle$. As can be seen, $\Phi(Z)$ is associated to a vector
$|\Phi\rangle$, which is a highest weight of the $N=3$ field.

Rather than work explicitly with (\ref{n3quant}), one could simply
consider what the infinitesimal transformations of the field are under
an infinitesimal superconformal map 
\begin{eqnarray}
(z, \theta_i)\mapsto (z+\delta z, \theta_i + \delta \theta_i)
\end{eqnarray}
This is useful to check closure as a lie algebra. It is useful to
introduce the quantity $\nu(z)
= \delta z + \theta_i\delta\theta_i$. The
transformation reads
\begin{eqnarray}\label{n3inf}
\delta\Phi(Z) &=& h(\partial_z\nu(Z))\Phi(Z) + \nu(Z)\partial_z\Phi(Z) +
\frac{1}{2}\sum_{j=1}^3(D_j\nu(Z))(D_j\Phi(Z)) +\nonumber\\&& \big((J_3D_1D_2 +
J_1D_2D_3 + J_2D_3D_1)(\nu(Z))\big)\Phi(Z)\\
&=& \frac{2h}{3}(D_1\delta\theta_1 + D_2\delta\theta_2 +
D_3\delta\theta_3)\Phi(Z) + (\delta z)\partial_z \Phi(Z) +\nonumber\\&&
\sum_{j=1}^3 (\delta\theta_j)\partial_{\theta_j}\Phi(Z) + \big(
(D_1\delta\theta_2-D_2\delta\theta_1)J_3 +
(D_2\delta\theta_3-D_3\delta\theta_2)J_1 +
\nonumber\\&&(D_3\delta\theta_1-D_1\delta\theta_3)J_2\big)\Phi(Z)\nonumber
\end{eqnarray}

The infinitesimal transformations form a Lie algebra, which can be
calculated explicitly from (\ref{n3inf}).
\begin{eqnarray}\label{n3lie}
& \lbrack \delta_{\nu_1}, \delta_{\nu_2} \rbrack \Phi(Z) =
\delta_{\nu_3}\Phi(Z) \nonumber\\
&\nu_3 = \nu_2(\partial_z\nu_1) - \nu_1(\partial_z\nu_2) +
\frac{1}{2}\sum_{i=1}^3 (D_i\nu_2)(D_i\nu_1) 
\end{eqnarray}
It is worth noting that the algebra closes if and only if the $J_i$
satisfy the commutation relations $\lbrack J_i, J_j \rbrack = -\frac{1}{2}\epsilon_{ijk}J_k$.

This can then be used to construct an ansatz for an OPE of
$\mathbb{T}(Z_1)\mathbb{T}(Z_2)$ (\ref{n3seope}), and then the modes calculated to
give the commutators of the quantum theory (\ref{n3comm}).
\begin{eqnarray}\label{n3seope}
\mathbb{T}(Z_1)\mathbb{T}(Z_2) &=& \frac{c}{Z_{12}} + 
\frac{\theta_{12,1}\theta_{12,2}\theta_{12,3}}{2Z_{12}^2}\mathbb{T}(Z_2)
+ \nonumber\\&&
\frac{\theta_{12,1}\theta_{12,2}\theta_{12,3}}{Z_{12}}\partial_w\mathbb{T}(Z_2)
+ \frac{\epsilon_{ijk}\theta_{12,i}\theta_{12,j}D_{2,k}}{4Z_{12}}\mathbb{T}(Z_2)
\end{eqnarray}
The first term gives rise to the central extension in the algebra, and
arises in precisely the same way as the bosonic case. This OPE shows
explicitly that $\mathbb{T}(Z)$ is a
weight $\frac{1}{2}$ field, although not primary. Since the
central charge does not appear for the super M\"obius subalgebra,
$\mathbb{T}$ can be thought of as a quasiprimary superfield, in the
trivial representation of $su(2)$. The modes of
this can then be calculated to give the $N=3$ algebra. Note that when
the classical algebra expressions appear in (\ref{n3quant}), there are
extra factors of $\frac{1}{2}$ appearing in (\ref{n3quant}). This
corresponds to the extra factors of $\frac{1}{2}$ appearing in
(\ref{n3comm}) when compared to the classical algebra.
\begin{eqnarray}\label{n3comm}
&\lbrack T_m^i, T_n^j \rbrack = -\frac{1}{2}\epsilon_{ijk}T_{m+n}^k +
mc\delta_{ij}\delta_{m+n,0} \quad \lbrack T_m^i, \psi_s \rbrack = 0 \nonumber\\&\nonumber\\& \lbrack
T_m^i, G_r^j \rbrack = \frac{1}{2}(\delta_{ij}m\psi_{r+m} -
\epsilon_{ijk}G_{r+m}^k) \quad \lbrack L_m, \psi_s \rbrack =
-(\frac{m}{2}+s)\psi_{m+s} \nonumber\\&\nonumber\\& \lbrack L_m, T_n^i \rbrack =
-nT_{m+n}^i \quad \lbrack G_r^i, \psi_s \rbrack = \frac{1}{2}T_{r+s}^i
\quad \lbrack L_m, G_r^i \rbrack = (\frac{m}{2}-r)G_{r+m}^i \nonumber\\&\nonumber\\&
\lbrack G_r^i, G_s^j \rbrack = \frac{1}{2}\delta_{ij}L_{r+s} +
\frac{1}{2}\epsilon_{ijk}(r-s)T_{r+s}^k - c(r^2 -
\frac{1}{4})\delta_{r+s,0}\delta_{ij}\nonumber\\&\nonumber\\& \lbrack L_m, L_n \rbrack =
(m-n)L_{m+n} - cm(m^2-1)\delta_{m+n,0} \quad \lbrack \psi_r, \psi_s
\rbrack = c\delta_{r+s,0}
\end{eqnarray}
which agrees with \cite{adem2}.

\section{The Neveu-Schwarz Algebra and its Verma Module}
The $N=3$ Neveu-Schwarz algebra is given by the above commutation relations
where $m \in \mathbb{Z}$, $r \in \mathbb{Z}+\frac{1}{2}$. The basis
can be changed so that the above relations are more useful
for representation theory. Consider a change of variables
\begin{eqnarray}
&T_m^+ = 2(iT_m^1 - T_m^2)\quad T_m^- = 2(iT_m^1 + T_m^2)\quad T_m^H =
-2iT_m^3 \nonumber\\&\nonumber\\ &G_r^+ = 4(G_r^2 - iG_r^1) \quad G_r^- = 4(G_r^2 +
iG_r^1) \quad G_r^H = 8iG_r^3 \quad k= -4c
\end{eqnarray}
Then, the commutation relations become, for $x \in \{H,
\pm\}$\cite{ft6}
\begin{eqnarray}\label{commrelns}
&\lbrack T_m^+, T_n^- \rbrack = 2T_{m+n}^H + 2km\delta_{m+n,0}\quad
\lbrack T_m^H, T_n^\pm \rbrack = \pm T_{m+n}^\pm\quad 
\lbrack T_m^\pm, T_n^\pm \rbrack = 0 \nonumber\\&\nonumber\\& \lbrack T_m^H, T_n^H
\rbrack = km\delta_{m+n,0} \quad\lbrack T_m^\pm, G_r^\pm \rbrack = 0
\quad \lbrack T_m^\mp, G_r^\pm \rbrack = -G_{r+m}^H \pm
8m\psi_{r+m}\nonumber\\&\nonumber\\& \lbrack T_m^\pm, G_r^H \rbrack=
-2G_{m+r}^\pm\quad \lbrack T_m^H, G_r^H \rbrack = 8m\psi_{m+r}\quad
\lbrack T_m^H, G_r^\pm \rbrack = \pm G_{r+m}^\pm\nonumber\\&\nonumber\\&
\lbrack \psi_s, G_r^\pm \rbrack = \mp T_{r+s}^\pm \quad \lbrack\psi_s,
G_r^H\rbrack = -2T_{r+s}^H \quad\lbrack G_r^\pm, G_s^\pm \rbrack = 0
\quad \lbrack T_m^x, \psi_s \rbrack = 0\nonumber\\&\nonumber\\& \lbrack
G_r^H, G_s^H \rbrack = -32L_{r+s} - 16k(r^2-\frac{1}{4})\delta_{r+s,0}
\quad \lbrack G_r^\pm, G_s^H \rbrack = 8(r-s)T_{r+s}^\pm\nonumber\\&\nonumber\\&
\lbrack G_r^+, G_s^- \rbrack = 16L_{r+s}
+8k(r^2-\frac{1}{4})\delta_{r+s,0} + 8(r-s)T_{r+s}^H
 \nonumber\\&\nonumber\\& \lbrack L_m, \psi_s \rbrack =
 -(\frac{m}{2}+s)\psi_{m+s} \quad \lbrack L_m, T_n^x \rbrack =
 -nT_{m+n}^x \quad \lbrack \psi_r, \psi_s
\rbrack = -\frac{k}{4}\delta_{r+s,0}\nonumber\\&\nonumber\\&
\lbrack L_m, G_r^x \rbrack =
 (\frac{m}{2}-r)G_{r+m}^x
\nonumber\\&\nonumber\\& \lbrack L_m, L_n \rbrack =
(m-n)L_{m+n} + \frac{k}{4}m(m^2-1)\delta_{m+n,0} \quad 
\end{eqnarray}
On the representations considered here, the algebra obeys
hermiticity conditions\cite{ft7}
\begin{eqnarray}
(\psi_r)^\dag = -\psi_{-r}\quad (T_m^+)^\dag = T_{-m}^- \quad (T_m^-)^\dag = T_{-m}^+\quad
(T_m^H)^\dag = T_{-m}^H\nonumber\\\nonumber\\
(G_r^H)^\dag = -G_{-r}^H \quad(L_n)^\dag = L_{-n}\quad (G_r^+)^\dag
= G_{-r}^-\quad (G_{r}^-)^\dag = G_{-r}^+
\end{eqnarray}
The highest weight conditions on a vector $\ket{\phi}$ are then
\begin{eqnarray}
T_m^x \ket{\phi} = 0 \quad G_r^x\ket{\phi} = 0 \quad L_m\ket{\phi} = 0
\quad\psi_r\ket{\phi}=0\quad T_0^+\ket{\phi} = 0
\end{eqnarray}
for $m,r>0$. The Cartan subalgebra is spanned by the elements $L_0,
T_0^H$, such that $L_0\ket{\phi} = h\ket{\phi}$, $T_0^H\ket{\phi} =
q\ket{\phi}$. The algebra of raising operators, i.e. the algebra
spanned by the elements giving the highest weight conditions, is
generated by $T_0^+$, $G_\frac{1}{2}^-$, $\psi_\frac{1}{2}$. Thus,
a vector $\ket{\chi}$ with the properties $T_0^+\ket{\chi}=0$,
$G_\frac{1}{2}^-\ket{\chi}=0$ and $\psi_\frac{1}{2}\ket{\chi}=0$ will
will obey the highest weight conditions. Consider the Verma module $V(h,q)$ for
a highest weight $\ket{\phi}$, with $L_0\ket{\phi} = h\ket{\phi}$,
$T_0^H\ket{\phi} = q\ket{\phi}$. A vector $\ket{\chi} \neq \ket{\phi}$
in the module defines a singular vector. The module itself admits a
decomposition
\begin{eqnarray}
V(h,q) = \bigoplus_{(m \geq 0)} \bigoplus_{(n \leq \frac{m}{2})} V_{m, n}
\end{eqnarray}
where $m \in \mathbb{Z}$ and $n \in \frac{\mathbb{Z}}{2}$. This can be
seen from the root structure
(\ref{commrelns}), and the highest weight conditions. 
An example of a singular vector occurs when $(h,q,k) =
(-\frac{1}{2},-1,k)$. Under such conditions, a singular
vector exists in $V_{\frac{1}{2},0}$.
\begin{eqnarray}\label{singvec1}
\ket{\chi} = T^-_0G^+_{-\frac{1}{2}}\ket{\phi}
\end{eqnarray}

\section{The Super M\"obius Group}
One might ask is how exactly does the theory of the M\"obius group
generalise. In the bosonic case, the lie algebra of the group can be
obtained by finding the globally defined vector fields on the Riemann
sphere. The Riemann sphere can be considered as a pair of complex
planes with transition function $w=\frac{1}{z}$ between them. In the
graded Riemann sphere case, one can choose a homomorphism between rings
of functions given by $(w,\chi_i)=(\frac{1}{z},
\frac{\theta_i\sqrt{-1}}{z})$. The south pole is $Z_s = (z, \theta_i)=(0,0,0,0)$, and the
north pole given by $Z_n = (\frac{1}{z}, \frac{\theta_i\sqrt{-1}}{z})=(0,0,0,0)$.
One can then ask what are the globally defined
graded vector fields. A basis of vector fields was calculated in
section 5. It can be seen that many of them are divergent at the
origin, or south pole. One must then check which vector fields are
well behaved at both poles. As an example, consider the vector field
(\ref{lnvf}). This is clearly divergent for $m<-1$ at the south
pole. To find out what $l_n$ looks like at the north pole, one uses
the techniques of graded one-forms and vector fields \cite{kos} to
find
\begin{eqnarray}
l_m = w^{-m+1}\frac{\partial}{\partial w} -
\frac{1}{2}(m-1)w^{-m}\chi_j\frac{\partial}{\partial \chi_j}
\end{eqnarray}
which is divergent for $m>1$ at the north pole. Thus, one can conclude
that $\{ l_1, l_0,
l_{-1}\} $ are globally defined. Similarly, one finds that the only other
globally defined vector fields are $\{ g^r_{\frac{1}{2}},
g^r_{-\frac{1}{2}}, t_0^i\} $. From the commutation relations,
(\ref{n3class}) it can be seen that the vector fields form a
closed subalgebra, namely $osp(3,2)$. One can then write down formal
group elements by exponentiation.
\begin{eqnarray}
&\exp({\lambda l_1}) : (z,\theta_i) \mapsto \frac{1}{1-\lambda z}(z,
\theta_i) \nonumber\\&
\exp({\lambda l_0}) : (z, \theta_i) \mapsto (e^\lambda z,
e^{\frac{\lambda}{2}}\theta_i)\nonumber\\&
\exp({\lambda l_{-1}}) : (z, \theta_i) \mapsto (z+\lambda, \theta_i)\nonumber\\&
\exp({\epsilon g_{-\frac{1}{2}}^j}) : (z,\theta_i) \mapsto (z- \epsilon \theta_j,
\theta_i, \theta_j + \epsilon) \quad i\neq j\nonumber\\&
\exp(\epsilon g_{\frac{1}{2}}^j) : (z,\theta_i) \mapsto
\frac{1}{1+\epsilon\theta_j}(z, \theta_i, \theta_j +
\epsilon z) \quad i\neq j\nonumber\\&
\exp(\lambda_i t_0^i) : (z, \theta_i) \mapsto (z,
M_{ij}(\lambda)\theta_j) \quad M_{ij}(\lambda) \in SO(3)
\end{eqnarray}
In particular, the $g_{-\frac{1}{2}}^i$ give supersymmetry generators,
the $g_{\frac{1}{2}}^i$ give special superconformal transformations, and
the $t_0^i$ give an R-symmetry. Writing these transformations as $Z
\mapsto Z'$, the corresponding transformations on the field become
\begin{eqnarray}
&e^{\lambda L_0}\Phi(Z)e^{-\lambda L_0} = e^{\lambda h}\Phi(Z') \qquad
e^{\lambda L_{-1}}\Phi(Z)e^{-\lambda L_{-1}} = \Phi(Z')\nonumber\\
&e^{\epsilon G_{-\frac{1}{2}}^i} \Phi(Z)e^{-\epsilon G_{-\frac{1}{2}}^i}
= \Phi(Z') \nonumber\\& e^{\rho G_{\frac{1}{2}}^i}\Phi(Z)e^{-\rho
G_{\frac{1}{2}}^i} =
\frac{1}{(1+\epsilon\theta_i)^h}e^{-2\rho\epsilon_{ijk}\theta_jJ_k}\Phi(Z')\nonumber\\&
e^{\lambda T_0^i}\Phi(Z)e^{-\lambda T_0^i} = e^{\lambda J_i}\Phi(Z')\nonumber\\&
e^{\lambda L_1}\Phi(Z)e^{-\lambda L_1} = \frac{1}{(1-\lambda
z)^h} e^{(\frac{\lambda}{(1-\lambda z)}\epsilon_{ijk}\theta_i\theta_j
J_k)} \Phi \big( \frac{z}{(1-\lambda
z)},\frac{\theta_i}{(1-\lambda z)}\big) 
\end{eqnarray}
Using these formal group elements, any
two points, $V = (v, \beta_i)$ and $U = (u, \alpha_i)$ say, can be mapped to
the north and south poles respectively. In a conformal field theory formalism, usually the
south pole is where the `in
vacuum' sits, and the north pole where the `out vacuum' sits. The
formal group element corresponding to this map is given by
\begin{eqnarray}\label{formgpelt}
(z, \theta_i) \mapsto (z', \theta_i') = \frac{\Big( z-u, \theta_i - \alpha_i +
\big(\frac{\alpha_i-\beta_i}{v-u}\big)(z-u)
\Big)}{( 1 + \frac{(\alpha-\beta)\cdot\theta}{v-u}) -
(1+\frac{\alpha\cdot\beta}{v-u})(\frac{z-u}{v-u})}
\end{eqnarray}
where $\alpha\cdot\beta = \sum_{i=1}^3\alpha_i\beta_i$.

To obtain this transformation, one can use $g_{-\frac{1}{2}}^i$  to
send $\alpha^i$ to $0$, $l_{-1}$ to move $u$ to $0$, $g_\frac{1}{2}$
to send $\beta_i$ to $0$ when $z=v$, and $l_1$ to send $v$ to
$\infty$. It is worth noting, that the only operators that have not
been used are the $t_0^i$ and $l_0$. This degree of freedom is
essentially a (complex) scale factor, and an $SO(3)$ action on the
$\theta$s. Thus, the even `co-ordinate' of the third point can be sent
anywhere one wishes, but on cannot quite do the same with the odd
`co-ordinates' of the third point. It should also be noted, that
this construction should generalise to $osp(N,2)$, i.e. with an
arbitrary number of odd co-ordinates.

The formal group element found (\ref{formgpelt}) implies that a correlation function of the form
\begin{eqnarray}\label{3ptfn}
\braket{0}{\Phi_1(V)\Phi_2(Z)\Phi_3(U)0}
\end{eqnarray}
can be superconformally mapped to a more typical presentation of the three-point function in conformal
field theory.
\begin{eqnarray}\label{3ptfnns}
\braket{0}{\Phi_1(\infty)\Phi_2(Z')\Phi_3(0)0} =\braket{\phi_1}{\Phi_2(Z')\phi_3}
\end{eqnarray}

\section{The Two-Point Function}
In conformal field theory it is known that global conformal invariance
is sufficient to solve for the two point function. This is indeed also
the case for $N=1,2$ \cite{west}. In this section, global $N=3$
invariance is used to solve for the two-point function. This
becomes quite a bit more complicated than smaller $N\leq 2$, calculationally, due to the presence of
non-abelian R-symmetry, manifested by the presence of $su(2)$
generators in the theory. More precisely, the primary fields are $\mathcal{A}_N\otimes \en\mathcal{H}
\otimes \mathcal{V}$ valued, where $\mathcal{V}$ is an $su(2)$ representation and
$\mathcal{H}$ is the Hilbert space that $|0\rangle$ belongs to. The super-Virasoro
operators are valued in $\en\mathcal{H}\otimes \en \mathcal{V}$.

The most convenient basis to work in is a
`charged' basis, where elements can be classified by their $su(2)$
charge, namely their $T^3_0$ eigenvalue. The basis is given by
\begin{eqnarray}
&\theta^+ = 2(i\theta_1-\theta_2)\quad & J^+ = 2(iJ_1-J_2)\nonumber\\
&\theta^- = 2(i\theta_1+\theta_2)\quad & J^- = 2(iJ_1+J_2)\nonumber\\
&\theta^H = i\theta_3\quad & J^H = -2iJ_3
\end{eqnarray}
The primary field $\Phi$ itself is the highest weight in an $su(2)$
representation, i.e. carries an $su(2)$ representation index, so that
\begin{eqnarray}
&J^+\Phi(Z) = (J^+)^A_{~B}\Phi^B(Z) = 0\nonumber\\
&J^H\Phi(Z) = (J^H)^A_{~B}\Phi^B(Z) = q\Phi^A(Z)
\end{eqnarray}
In the following, $\Phi_i(Z)$ has conformal weight $h_i$ and spin $q_i$.
The action of the twelve globally defined generators on $\Phi(Z)$ can then be
given in Lie algebra form. The infinitesimal transformations are
\begin{eqnarray}
&\lbrack L_{-1}, \Phi \rbrack = \partial_z \Phi
\qquad\lbrack G_{-\frac{1}{2}}^\pm , \Phi \rbrack = \pm (\theta^\pm
\partial_z + 8\partial_{\theta^\mp})\Phi\nonumber\\&\nonumber\\&
\lbrack G_{-\frac{1}{2}}^H, \Phi \rbrack = -4(\theta^H\partial_z +
\partial_{\theta^H})\Phi\qquad
\lbrack T_0^H, \Phi \rbrack = (\theta^-\partial_{\theta^-} -
\theta^+\partial_{\theta^+} + J^H)\Phi  \nonumber\\&\nonumber\\&
\lbrack L_0, \Phi \rbrack = \big(h + z\partial_z +
\frac{1}{2}(\theta^+\partial_{\theta^+} + \theta^-\partial_{\theta^-}
+ \theta^H\partial_{\theta^H} )\big)\Phi\nonumber\\&\nonumber\\&
\lbrack T^\pm_0, \Phi \rbrack =
(\mp\frac{1}{2}\theta^\pm\partial_{\theta_H} \pm
4\theta^H\partial_{\theta^\mp} + J^\pm)\Phi\nonumber\\&\nonumber\\&
\lbrack L_1, \Phi \rbrack = \big(2hz + z(z\partial_z +
\theta^+\partial_{\theta^+} + \theta^-\partial_{\theta^-}+
\theta^H\partial_{\theta^H}) + \frac{1}{8}\theta^+\theta^-J^H\nonumber\\&\qquad\qquad
+\frac{1}{4}\theta^+\theta^HJ^-
-\frac{1}{4}\theta^-\theta^HJ^+\big)\Phi\nonumber\\&\nonumber\\&
\lbrack G^\pm_{\frac{1}{2}}, \Phi \rbrack = (\pm2h\theta^\pm \pm
\theta^\pm z\partial_z \pm 8z\partial_{\theta^\mp}
\pm\theta^\pm\theta^H\partial_{\theta^H} +
\theta^+\theta^-\partial_{\theta^\mp} \nonumber\\&\qquad\qquad
+ 2\theta^H J^\pm + \theta^\pm J^H)\Phi\nonumber\\&\nonumber\\&
\lbrack G^H_\frac{1}{2}, \Phi \rbrack = (-8h\theta^H - 4\theta^H
z\partial_z - 4z\partial_{\theta^H} -
4\theta^H\theta^-\partial_{\theta^-} -
4\theta^H\theta^+\partial_{\theta^+}\nonumber\\&\qquad\qquad + \theta^-J^+ - \theta^+J^-)\Phi
\end{eqnarray}
Note that under the $T_0^H$ operator, $\theta^+$ and $\theta^-$ are
`charged', i.e. they possess a non-zero $T^H_0$ eigenvalue. The two
point function, $\braket{0}{\Phi_1(Z_1)\Phi_2(Z_2)0} = \langle \Phi_1(Z_1)\Phi_2(Z_2)\rangle$ is, as
a function, a function of $Z_1 = (z, \theta_i)$ and $Z_2 =
(w,\chi_i)$. Since the $\Phi_i$ are also highest weight vectors of
$su(2)$ representations, $\mathcal{V}_i$, the two point function is an
element of $\mathcal{V}_1\otimes \mathcal{V}_2$. The $L_{-1}$ condition on the two-point
function reads
\begin{eqnarray}\label{l-1cndn}
\mathcal{L}_{-1} \langle \Phi_1(Z_1)\Phi_2(Z_2)\rangle = (\partial_z +
\partial_w)\langle \Phi_1(Z_1)\Phi_2(Z_2)\rangle = 0
\end{eqnarray}
implying that $\langle \Phi_1(Z_1)\Phi_2(Z_2)\rangle$ is a function of
$(z-w)$ and $\theta_i, \chi_i$. Applying the $G_{-\frac{1}{2}}^x$
conditions yields similar equations to (\ref{l-1cndn}). These
conditions show that $\langle
\Phi_1(Z_1)\Phi_2(Z_2)\rangle$ is a function of
\begin{eqnarray}
(\theta^--\chi^-),\quad (\theta^H-\chi^H),\quad
(\theta^+-\chi^+),\nonumber\\ s=(z-w+\frac{1}{8}(\theta^-\chi^+ +
\theta^+\chi^-) + \theta^H\chi^H)
\end{eqnarray}
The $L_0$ condition gives a scaling condition, from which the most
general form of the two point function can be seen to be
\begin{eqnarray}
\langle \Phi_1(Z_1)\Phi_2(Z_2)\rangle = &\frac{a}{s^{h_1+h_2}} +
\frac{\epsilon_+(\theta^+-\chi^+)}{s^{h_1+h_2+\frac{1}{2}}} +
\textrm{(two similar terms)}\nonumber\\&
+\frac{b_{+H}(\theta^+-\chi^+)(\theta^H-\chi^H)}{s^{h_1+h_2+1}} +
\textrm{(two similar terms)}\nonumber\\&+ \frac{\eta
(\theta^+-\chi^+)(\theta^--\chi^-)(\theta^H-\chi^H)}{s^{h_1+h_2+\frac{3}{2}}}
\end{eqnarray}
The $T^H_0$ condition includes $su(2)$ elements. It is worth writing
this condition out explicitly, to illustrate the action of the
elements. Putting in all the tensor products between $su(2)$
representations explicitly, the condition reads
\begin{eqnarray}
\Big( (\mathbb{I}\otimes\mathbb{I})(\theta^-\partial_{\theta^-} -
\theta^+\partial_{\theta^+} + \chi^-\partial_{\chi^-} -
\chi^+\partial_{\chi^+}) + \nonumber\\J^H\otimes\mathbb{I} +
\mathbb{I}\otimes J^H\Big)\langle \Phi_1(Z_1)\otimes\Phi_2(Z_2)\rangle=0
\end{eqnarray}
where
\begin{eqnarray}\label{jhaction}
(J^H\otimes\mathbb{I} + \mathbb{I}\otimes J^H)\langle
\Phi_1(Z_1)\otimes\Phi_2(Z_2)\rangle = \langle
(J^H\Phi_1)(Z_1)\otimes\Phi_2(Z_2)\rangle +\nonumber\\ \langle
\Phi_1(Z_1)\otimes(J^H\Phi_2)(Z_2)\rangle = (q_1+q_2)\langle
\Phi_1(Z_1)\otimes\Phi_2(Z_2)\rangle
\end{eqnarray}
This condition gives three possible cases
\begin{eqnarray}
&&q_1+q_2 = 0 \Rightarrow \textrm{only (}a, \epsilon_H, b_{+-},
\eta\textrm{) non-zero}\nonumber\\
&&q_1+q_2 = 1 \Rightarrow \textrm{only (}\epsilon_+, b_{+H}\textrm{) non-zero}\nonumber\\
&&q_1+q_2 = -1 \Rightarrow \textrm{only (}\epsilon_-, b_{-H}\textrm{) non-zero}
\end{eqnarray}
Replacing $H$ with $+$ in (\ref{jhaction}), it can be seen that  $J^+\otimes\mathbb{I} + \mathbb{I}\otimes J^+$ annihilates $\langle
\Phi_1(Z_1)\otimes\Phi_2(Z_2)\rangle$. The $T_0^+$ condition then
gives - if $q_1+q_2=-1$, then $\epsilon_-,b_{-H}=0$ - if $q_1+q_2=0$, then
$\epsilon_+, b_{+H}=0$ - and gives no extra conditions if
$q_1+q_2=1$. Thus, the $q_1+q_2=-1$ case is irrelevant.
$J^-$ is an operator that can cause calculational difficulties. The $T_0^-$
condition can be used to relate $\langle (J^-\Phi_1)\Phi_2 \rangle$
and $\langle \Phi_1(J^-\Phi_2) \rangle$.

Consider now the $L_1$ condition. This contains a term like 
\begin{eqnarray}
\theta^+\theta^H\langle (J^-\Phi_1)\Phi_2\rangle + \chi^+\chi^H\langle \Phi_1(J^-\Phi_2)\rangle\nonumber
\end{eqnarray}
The $T_0^-$ condition can be used to relate this to a term of the form
\begin{eqnarray}
(\theta^+\theta^H-\chi^+\chi^H)\langle (J^-\Phi_1)\Phi_2\rangle \nonumber
\end{eqnarray}
Thus the condition implies that all those terms that cannot be
factored by $(\theta^+\theta^H-\chi^+\chi^H)$ are zero. A similar
condition arises for the $G_{\frac{1}{2}}^x$ conditions. After much
tedious algebra, one finds that
\begin{eqnarray}\label{2ptfn}
\langle\Phi_1(Z_1)\Phi_2(Z_2)\rangle =\left\{\begin{array}{ll} 
\frac{a}{s^{h_1+h_2}}
&\textrm{if}\quad h_1=h_2,\quad q_1=q_2=0\\
\frac{b_{+H}(\theta^+-\chi^+)(\theta^H-\chi^H)}{s^{h_1+h_2+1}}&\textrm{if}\quad
h_1=h_2\quad q_1+q_2=1,\\& \quad q_1,q_2\neq 0\\
0 &\textrm{otherwise}
\end{array}\right.
\end{eqnarray}
This has important applications to fusion. Considering the three-point
function 
\begin{eqnarray}
\mathcal{F}_{123} =
\langle\Phi_1(Z_1)\Phi_2(Z_2)\Phi_3(Z_3)\rangle\nonumber
\end{eqnarray}
where $\Phi_1(Z_1)$ is a `probe field' i.e. one can choose its $(h,q)$
parameters, call them $(h_1,q_1)$. $\Phi_2(Z_2)$ and
$\Phi_3(Z_3)$ will have an OPE, which schematically looks like (i.e. omitting pole
structure and other factors)
\begin{eqnarray}\label{schemfus}
\Phi_2(Z_2)\Phi_3(Z_3) \sim \sum_n\Psi_n(Z_3)
\end{eqnarray}
that may be unknown, namely one may not know the $(h,q)$ of the $\Psi_n$. One can ask if the OPE between $\Phi_2(Z_2)$ and
$\Phi_3(Z_3)$ can be deduced if one knows the values of
$\mathcal{F}_{123}$, for all $h_1$ and $q_1$. From (\ref{2ptfn}), one
can see that for $\langle \Phi_1(Z_1), \Psi_n(Z_3)\rangle$ to be
non-zero, a unique $(h_1, q_1)$ must be chosen. This choice determines the
$(h,q)$ of $\Psi_n(Z_3)$. Thus one can make the statement that
knowing when the three point function $\mathcal{F}_{123}$ vanishes is equivalent to
knowing what $\Psi_n(Z_3)$ are in (\ref{schemfus}). These then give rise to
the fusion rules. 

One should note that global conformal invariance of the
theory almost fixes three super co-ordinates, as can be seen from
(\ref{formgpelt}). In the mapping from (\ref{3ptfn}) to
(\ref{3ptfnns}) one can map $V$ and $U$ to the north and south
poles. One can also map $z'$ from $Z'=(z',\theta'_i)$ in
(\ref{3ptfnns}), to wherever desired, using $L_0$. There is not enough freedom to
move the $\theta'_i$ wherever desired. Thus, one would expect that the three-point function
could also be computed, up to an arbitrary function in
$\theta'_i$. After expanding this function into components,
this can be seen as being computable up to some arbitrary constants.

\section{Other Applications}
This section is strictly speaking a list of things that could be done,
in the $N=3$ theory. Since most of these things are very calculationally
intensive, the author has not checked the details.

An interesting question is analysing the constraints that singular
vectors give on three-point functions. If $\ket{\chi}$ is a singular
vector in a module with highest weight $\ket{\phi_1}$, then what does
the requirement that
\begin{eqnarray}
\braket{\phi_3}{\Phi(Z)\chi}\nonumber
\end{eqnarray}
vanish imply about
\begin{eqnarray}
\braket{\phi_3}{\Phi(Z)\phi_1}\nonumber
\end{eqnarray}
Algebraically, this is in fact quite
difficult, and the author has not managed to accomplish this. The main
complication is that the composition between primary fields in the
correlator are a tensor product between $su(2)$ representations $\mathcal{V}$
and $\mathcal{V}'$ and an $\en\mathcal{H}$
composition. This means that the only super-Virasoro operators that can be
transferred across the tensor product are those that have $\en\mathcal{V}$ part
proportional to the identity. The most obvious case where this applies
is where all the fields have $q=0$, i.e. they are all $su(2)$
singlets. Following \cite{dor2}, the lowering operators
acting appearing in $\ket{\chi}$ can be rewritten in terms of operators
that have commutator with a primary field in $(z, \theta_i)$ given by
a polynomial in $(z, \theta_i)$, namely
\begin{eqnarray}
\mathcal{L}_m &=& -L_m + \frac{1}{z}L_{m+1} +
\frac{1}{16z}(\theta^+G^-_{m+\frac{1}{2}} -
\theta^-G^+_{m+\frac{1}{2}} + 2\theta^H G^H_{m+\frac{1}{2}}) +\nonumber\\
&&\frac{m+1}{4z}(\theta^-\theta^HT_m^+ - \theta^+\theta^HT_m^- -
\theta^+\theta^-T_m^H)\nonumber\\
\mathcal{G}_r^\pm &=& -G_r^\pm + \frac{1}{z}G_{r+1}^\pm -
\frac{2\theta^H}{z}T_{r+\frac{1}{2}}^\pm - \frac{\theta^\pm}{z}T_{r+\frac{1}{2}}^H\nonumber\\
\mathcal{G}_r^H &=& -G_r^H + \frac{1}{z}G_{r+1}^H -
\frac{\theta^-}{z}T_{r+\frac{1}{2}}^+ + \frac{\theta^+}{z}T_{r+\frac{1}{2}}^-\nonumber\\
\mathcal{T}_m^\pm &=& -T_m^\pm + \frac{1}{z}T_{m+1}^\pm - \frac{\theta^\pm}{z}\psi_{m+\frac{1}{2}}\nonumber\\ 
\mathcal{T}_m^H &=& -T_m^H + \frac{1}{z}T_{m+1}^H + \frac{2\theta^H}{z}\psi_{m+\frac{1}{2}}\nonumber\\ 
\mathcal{P}_r &=& -\psi_r + \frac{1}{z}\psi_{r+1}
\end{eqnarray}
One then finds 
\begin{eqnarray}
&\lbrack \mathcal{L}_m, \Phi(Z) \rbrack = hz^m, \qquad \lbrack
\mathcal{G}_r^\pm, \Phi(Z) \rbrack = \pm 2h\theta^\pm
z^{r-\frac{1}{2}}\nonumber\\ &\qquad \lbrack\mathcal{G}_r^H,
\Phi(Z)\rbrack = -8h\theta^Hz^{r-\frac{1}{2}}, \qquad
\lbrack\mathcal{T}_m^\pm, \Phi(Z)\rbrack = \pm
h\theta^\pm\theta^Hz^{m-1}, \nonumber\\ &\lbrack\mathcal{T}_m^H, \Phi(Z)\rbrack = -\frac{1}{4}
h\theta^+\theta^-z^{m-1},\qquad \lbrack \mathcal{P}_r, \Phi(Z)\rbrack
= \frac{1}{8}h\theta^+\theta^-\theta^Hz^{r-\frac{3}{2}}
\end{eqnarray}
Now one can commute the operators from $\ket{\chi}$ past $\Phi$
without introducing differential operators. The operators
$\mathcal{L}_m$ etc may now be re-expanded in terms of $L_m$
etc. Some of these operators will annihilate $\bra{\phi_3}$. Those
that do not, and are not diagonal, must be processed using the descent
equations on $\ket{\Phi(Z)\phi_1}$ \cite{bpz}. This then yields a
set of polynomial equations giving conditions on the weights of the
primary fields.

From a differential equation point of
view, the question of singular vectors may not be such a difficult problem. As in \cite{bpz}, the
lowering operators can be written as contour integrals, e.g.
\begin{eqnarray}\label{contintops}
&&L_{-k}(z) = \frac{1}{2\pi i}\oint dw L(z) (w-z)^{1-k} \nonumber\\ &&G_{-r}(z)  =
\frac{1}{2\pi i}\oint dw G(z) (w-z)^{\frac{1}{2}-r}\qquad\textrm{etc..}
\end{eqnarray}
The condition of a singular vector, $\mathcal{N}\ket{\phi_3}$, where
$\mathcal{N}$ are some lowering operators, can be written as 
\begin{eqnarray}
\langle \Phi_1(Z_1)\Phi_2(Z_2) (\mathcal{N}\Phi_3)(Z_3) \rangle = 0\nonumber
\end{eqnarray}
This then gives rise, via (\ref{contintops}) and the OPE, to a
differential equation on the three point function. One would expect the
three point function  to look like a product of powers of differences,
as in the case of the two point function, e.g. $s,
(\theta^+-\chi^+)$. As in the bosonic case, this should give rise to a
polynomial in the $h_i, q_i$ of the fields concerned. The difference now, is that
the presence of $J^-$ operators will give independent equations,
e.g. $\mathcal{F}_{123}\in
\mathcal{V}_1\otimes\mathcal{V}_2\otimes\mathcal{V}_3$, hence
$(\mathbb{I}\otimes\mathbb{I}\otimes J^-)\mathcal{F}$ is linearly independent
of $(\mathbb{I}\otimes\mathbb{I}\otimes\mathbb{I})\mathcal{F}$. Other
than this, the calculations should proceed precisely as in the bosonic
case.

\section{Conclusions}
Starting from a graded Riemann sphere, a superconformal field theory
was constructed. The construction roughly parallels that of the
bosonic case, namely defining sections of a line bundle on a Riemann
sphere, and rewriting the infinitesimal transformations of these
sections as operator product expansions. Two ways were used to
introduce a Ramond field, one by introducing a branch cut, the other
by altering the preserved one-form. This suggests that looking at
various sheaves on a graded Riemann sphere may be a potentially useful
way of realizing fields in a superconformal field theory.

The super OPEs, together with
an understanding of how the symmetries act on the graded Riemann
sphere, were sufficient to compute the $N=3$ two-point function, up to
multiplicative constants. In addition, it was illustrated how, in principle, the $N=3$ three point function and
conditions given by singular vectors on the three point function could
be calculated. It should be pointed out that the method of calculation
was entirely in superfield formalism, and hence manifestly supersymmetric.

The only case that has really been studied here is the $N=3$ case,
based on a Riemann Surface of genus zero. How this generalises to
higher genus is an interesting question. An even more interesting
question, is the $N=4$ case. Processing the $N=4$ theory through this
machinery, does not produce the full OPE of the theory. There is a
log term missing from the OPE corresponding to a $U(1)$
charge. A question then arises, how to extend the framework of a
graded Riemann sphere to incorporate this log term. Many parts of the $N=4$
theory will in fact look like the $N=3$ theory, since the $N=4$
currents arising from R-symmetry are a pair of commuting $su(2)$
currents.

The author would like to thank Dr M. D\"orrzapf, Dr J. Evans,
J. Lucietti, Prof N. Manton, Prof H. Osborn and Prof B. Totaro for helpful discussions. The author is
funded by PPARC.


\newpage
\begin{thebibliography}{99}
\bibitem{adem} Ademollo, M., et al ``Supersymmetric Strings and
Colour Confinement,'' Phys. Lett. \textbf{62B}, 105-110 (1976)
\bibitem{adem2} Ademollo, M., et al ``Dual String Models with
Non-Abelian Colour and Flavour Symmetries,'' Nuc. Phys.
\textbf{B114}, 297-316 (1976)
\bibitem{ali} Ali, A. ``Classification of Two Dimensional $N=4$
Superconformal Symmetries,'' hep-th/9906096
\bibitem{bat1} Batchelor, M., ``Two approaches to Supermanifolds,''
Trans. Amer. Math. Soc. \textbf{253}, 329-338 (1979)
\bibitem{bat2} Batchelor, M., ``Graded Manifolds and Supermanifolds,''
Mathematical Aspects of Superspace  NATO ASI Series, Math and
Phys Sciences \textbf{132}, 91-133 (1983)
\bibitem{bpz} Belavin, A., Polyakov, A., Zamolodchikov, A.,
  ``Infinite Conformal Symmetry in two-dimensional Quantum Field
  Theory,'' Nuc. Phys. \textbf{B241}, 333-380 (1984)
\bibitem{brem} Bremner, M., ``Superconformal extensions of the
Witt algebra,'' MSRI 03309-90 Berkeley University, (1990)
\bibitem{ck} Chang, D., Kumar, A., ``Representations of $N=3$
  Superconformal algebra,'' Phys. Lett. \textbf{B193}, 181-186 (1987)
\bibitem{cohn} Cohn, J., ``$N=2$ Super-Riemann Surfaces''
Nuc. Phys. \textbf{B284}, 349-364 (1987)
\bibitem{dew} DeWitt, B., ``Supermanifolds,'' Cambridge Unversity Press (1992)
\bibitem{dor2} D\"orrzapf, M., ``Singular Vectors of the $N=2$
  Superconformal Algebra,''
Int. J. Mod. Phys. A \textbf{10}, 2143-2180 (1995)
\bibitem{dor1} D\"orrzapf, M., ``The definition of Neveu-Schwarz
superconformal fields and superconformal transformations,''
Rev. Math. Phys \textbf{11}, 137-169 (1999)
\bibitem{difr} Di Francesco, P., Mathieu, P., Senechal, D.,
``Conformal Field Theory,'' Springer-Verlag (1996)
\bibitem{gidd} Giddings, S., ``A Brief Introduction to Superriemann
Surface Theory,'' Trieste school (1988)
\bibitem{gsrs} Giddings, S., Nelson, P., ``The Geometry of Super
  Riemann Surfaces,'' C.M.P. \textbf{116}, 607-634 (1988)
\bibitem{gins} Ginsparg, P., ``Applied Conformal Field Theory,'' Les
Houches (1988)
\bibitem{godd1} Goddard, P., ``Meromorphic conformal field
theory,'' Adv. Ser. Math. Phys. 7, Infinite-dimensional Lie algebras
and groups 556-587, World Scientific Publishing (1988)
\bibitem{bred} Hartshorne, R., ``Algebraic Geometry,'' Springer-Verlag
  (1977)
\bibitem{hodg} Hodgkin, L. ``Super-Beltrami Differetials,''
  Class. Quantum Grav. \textbf{6}, 1725-1738 (1989)
\bibitem{kac2} Kac, V., ``Lie Superalgebras,'' Adv. Maths. 
\textbf{26}, 8-96 (1977)
\bibitem{kac3} Kac, V., van de Leur, J., ``On classification of
superconformal algebras,'' Proceedings, Strings (1988)
\bibitem{kac} Kac, V., ``Vertex Algebras for Beginners,'' AMS (1997)
\bibitem{kos} Kostant, B., ``Graded Manifolds, Graded Lie Theory,
and Prequantization,'' Differential geometry in mathematical
physics. Lecture Notes in Mathematics, Vol 570
\bibitem{leit} Leites, D., ``Introduction to the Theory of
Supermanifolds,'' Russian Math Surveys \textbf{35:1}, 1-64 (1980)
\bibitem{lust} L\"ust, D., Theisen, S., ``Lectures on String Theory,''
Springer (1989)
\bibitem{rog1} Rogers, A., ``A Global Theory of Supermanifolds'' J.
Math. Phys. \textbf{21}(6), 1352-1365 (1980)
\bibitem{rog2} Rogers, A., ``Consistent superspace integration,'' J.
Math. Phys. \textbf{26}(3), 385-392 (1985)
\bibitem{rog3} Rogers, A., ``Graded Manifolds, Supermanifolds and
Infinite-Dimensional Grassmann Algebras,'' C.M.P. \textbf{105},
375-384 (1986)
\bibitem{ks} Schoutens, K., ``$O(N)$-extended Superconformal Field
Theory in Superspace,'' Nuc. Phys. \textbf{B295}, 634-652 (1988)
\bibitem{gw} Watts G., ``Null vectors of the superconfomal algebra:
The Ramond sector,'' Nuc. Phys. \textbf{B407}, 213-236 (1993)
\bibitem{west} West, P., ``Introduction to Supersymmetry and
Supergravity'' World Scientific (1990)
\bibitem{ft1}Assume otherwise, so that $\kappa\kappa ' = 1 = \kappa
'\kappa$, and $\kappa\kappa ''=1=\kappa ''\kappa$. Subtracting and
factorising gives $(\kappa '-\kappa '')\kappa = 0 = \kappa(\kappa
'-\kappa '')$. Using either inverse on these equations shows $\kappa '
- \kappa '' = 0$.
\bibitem{ft2}The commutators are graded, i.e. $\lbrack A, B \rbrack =
  AB - (-1)^{p(A)p(B)}BA$
\bibitem{ft3}$\epsilon_{123}=1$ and $\epsilon$ is
antisymmetric in all its indices. Summation convention is used over
repeated indices
\bibitem{ft4}The commutation
relations are $\lbrack J_i, J_j \rbrack =
-\frac{1}{2}\epsilon_{ijk}J_k$. These are precisely the commutation
relations of the sub-algebra formed by the $\frac{1}{2}t_0^i$
\bibitem{ft5}In the bosonic
case, a vertex operator is characterised uniquely by its action on a
vacuum, and can be defined by $\phi(z)|0\rangle =
exp(zL_{-1})|\phi\rangle$. On a super-complex plane, this can be generalised
to $\Phi(Z)|0\rangle=exp(zL_{-1})exp(\theta_1
G_{-\frac{1}{2}}^1)\ldots exp(\theta_N
G_{-\frac{1}{2}}^N)|\Phi\rangle$, giving the operator associated to a
state.
\bibitem{ft6}The author thanks M. D\"orrzapf for checking these relations
\bibitem{ft7}More precisely, these come from an
ungraded involution on the algebra. On any representation of the
algebra that admits a hermitian contragredient form, this involution
then gives the adjoint with respect to that form.
\end{thebibliography}
\end{document}